\documentclass[5p]{elsarticle}
\usepackage{multirow}
\usepackage{float}
\usepackage[hidelinks]{hyperref}
\usepackage{subcaption}
\usepackage{lipsum}
\usepackage{placeins}
\usepackage{amstext}

\usepackage{graphicx}

\begin{document}

\begin{frontmatter}

\title{Quantifying Noise Limitations of Neural Network Segmentations in
High-Resolution Transmission Electron Microscopy.}

\author[camd]{Matthew Helmi Leth Larsen}
\author[nlab]{William Bang Lomholdt}
\author[camd]{Cuauhtemoc Nu$\tilde{\text{n}}$ez Valencia}
\author[nlab]{Thomas W. Hansen}
\author[camd]{Jakob Schiøtz}

\address[camd]{Computational Atomic-scale Materials Design (CAMD), Department of Physics,
  Technical University of Denmark, DK-2800 Kgs.\  Lyngby, Denmark}
\address[nlab]{National Center for Nano Fabrication and Characterization,
  Technical University of Denmark, DK-2800 Kgs.\  Lyngby, Denmark}

\begin{abstract}
Motivated by the need for low electron dose transmission electron microscopy imaging, we report the optimal frame dose (\emph{i.e.} $e^-/\text{\AA}^{2}$) range for object detection and segmentation tasks with neural networks. The MSD-net architecture shows promising abilities over the industry standard U-net architecture in generalising to frame doses below the range included in the training set, for both simulated and experimental images. It also presents a heightened ability to learn from lower dose images. The MSD-net displays mild visibility of a Au nanoparticle at 20-30 $e^-/\text{\AA}^{2}$, and converges at 200 $e^-/\text{\AA}^{2}$ where a full segmentation of the nanoparticle is achieved. Between 30 and 200 $e^-/\text{\AA}^{2}$ object detection applications are still possible. This work also highlights the importance of modelling the modulation transfer function when training with simulated images for applications on images acquired with scintillator based detectors such as the Gatan Oneview camera. A parametric form of the modulation transfer function is applied with varying ranges of parameters, and the effects on low electron dose segmentation is presented.
\end{abstract}

%%%%%%%%%%%%%%%%%%%%%%%%%%%%%%%%%%%%%%%%%%%%%%
%%                                          %%
%% The keywords begin here                  %%
%%                                          %%
%% Put each keyword in separate \kwd{}.     %%
%%                                          %%
%%%%%%%%%%%%%%%%%%%%%%%%%%%%%%%%%%%%%%%%%%%%%%


\end{frontmatter}

%\linenumbers

\section{Introduction}
\label{sec:introduction}

High-resolution transmission electron microscopy (HR-TEM) is a primary method to characterise materials at the atomic scale, and is a method where an abundance of data can be obtained. HR-TEM can provide a greater temporal resolution, as opposed to scanning transmission electron microscopy, by illuminating the entire sample simultaneously. It does this, however, at the cost of the signal-to-noise ratio (SNR). Increasing the frame dose of the image can assist in increasing the interpretability of each image, but it is usually undesired to do so to avoid electron beam induced effects that are yet to be completely understood \cite{Batson2008Beameffect,VanDyck:2015fg,Egerton2019RadiationTEM}. The averaging of many images is another method to increase the SNR. This method, however, lowers the temporal resolution and can be a tedious task that requires complex image alignment to ensure sensible results. Image alignment is also frame dose dependent as the alignment between low SNR images becomes difficult. Addressing SNR related issues has led to a quest in developing denoising methods \cite{Du2015AMicrographs,Furnival2017DenoisingThresholding}, often using neural networks \cite{Lin2021:TEMImageNet,Vincent2021DevelopingSignal-to-Noise}.

The field has seen a steady increase in applying machine learning solutions to solve various tasks of analysing and interpreting data. The importance of including machine learning in a standard workflow for HR-TEM characterisation is highlighted in multiple works \cite{Spurgeon2021TowardsMicroscopy, Treder2022ApplicationsMicroscopy}. Pipelines for training neural networks for segmentation with hand labelled experimental data have been developed, such as the pipeline by Groschner \emph{et al.} \cite{Groschner2021MachineData}, where segmented nanoparticles are classified to acquire sufficient statistics on various classes of nanoparticles. The segmentation of nanoparticles is also useful for tracking dynamic behaviour across frames \cite{Faraz2022DeepStudies}. Simulated images provide possibilities for large amounts of accurately labelled images for training and have been used to identify and analyse atomic columns in experimental HR-TEM sequences \cite{Madsen2018AImages, Ragone2020AtomicLearning}. Vincent \emph{et al.} also used simulated images to train neural networks for denoising \cite{Vincent2021DevelopingSignal-to-Noise}. Due to the successes of deep learning powered analysis of HR-TEM images many are seeking to understand the important aspects of optimising the generalisability and applications of deep learning models, and attempt to determine the best models available in comparison to eachother and other thresholding or clustering methods \cite{Horwath2020UnderstandingImages, Saaim2022InSegmentation}.

In this work we look to continue the search for optimal simulated training data and neural networks, by considering the specific task of low SNR segmentation. We will report the optimal frame dose range for reliable segmentations by neural networks and how to control that range by tuning simulations. For this task we focus on the segmentation of metallic nanoparticles, more specifically CeO$_2$ supported Au nanoparticles. Au nanoparticles serve as a useful system for studying catalytic properties of metallic nanoparticles \cite{Liu2019}. 

\begin{figure*}[!t]
    \centering
    \includegraphics[width=0.7\linewidth]{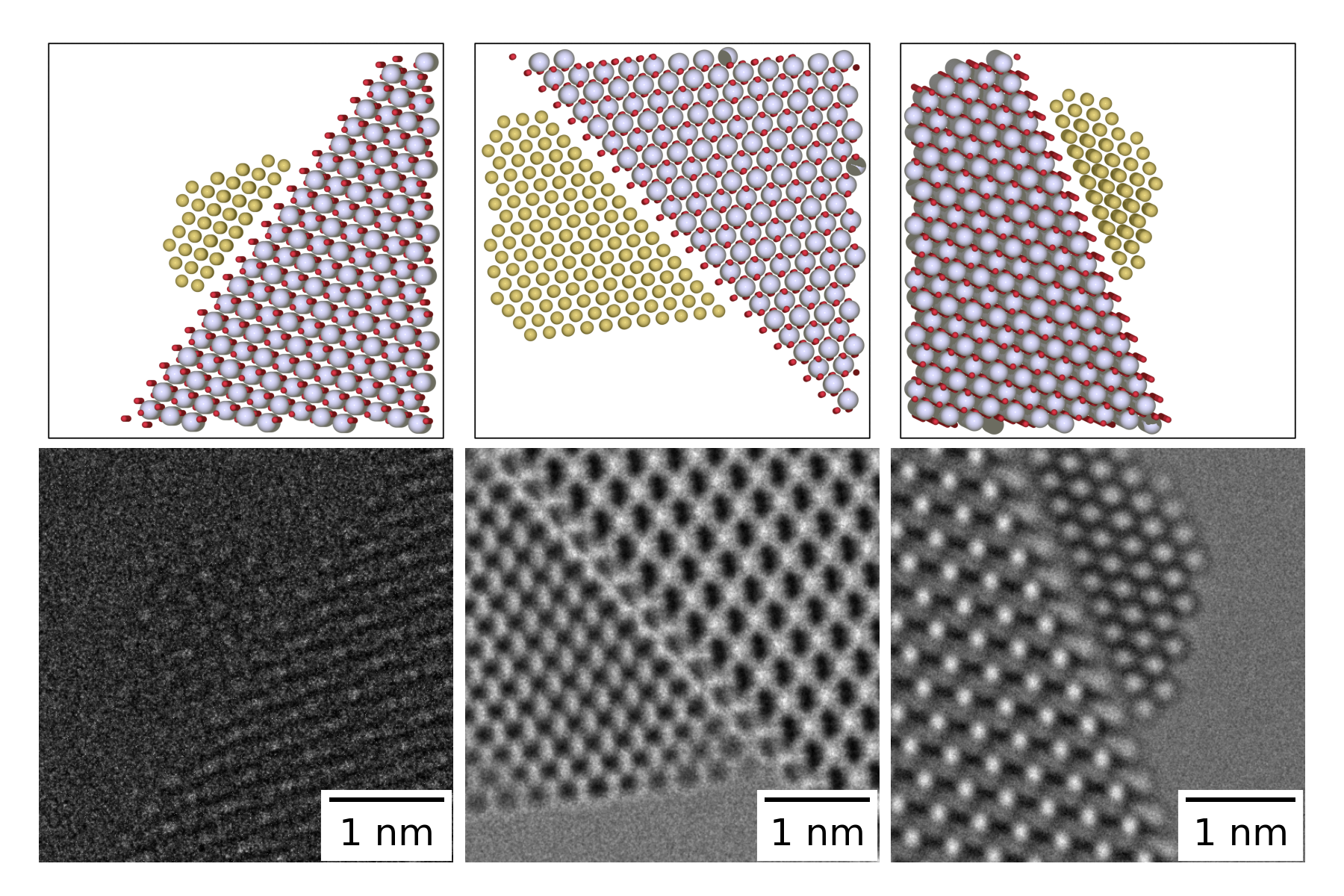}
    \caption{Top: A 3D rendered visual of three random CeO$_2$ supported Au nanoparticle structures constructed with ASE and Bottom: A corresponding HR-TEM image simulated with abTEM using random microscope parameters from Table \ref{tab:parameters}.}
    \label{fig:Au_visual}
\end{figure*}

\section{Methods}
\label{sec:methods}

This work utilises two different network architectures to perform the segmentations. The first is the well recognised U-net \cite{Ronneberger:2015gk} architecture used by Madsen \emph{et al.} \cite{Madsen2018AImages}, with the only modification that the skip connections are now concatenations rather than element-wise additions \cite{LethLarsen2022ReconstructingLearning}. The other is the MSD-net introduced by Pelt \emph{et al.} \cite{Pelt2017}, which was highlighted to be a robust network against low SNR images. Both networks are convolutional neural networks that differ in their approach to capturing information from varying spatial distances in the image. The U-net's approach gives its characteristic U-shaped architecture due to the numerous down-sampling and up-sampling layers to spread the convolutional kernel over patterns with large spatial coverage. The MSD-net in contrast maintains the same resolution throughout all layers, but dilates the convolutional kernel to spread the weights over a larger spatial region. Specific hyper-parameters regarding the network architectures can be seen in the Supplementary Online Information (SOI) together with SOI Fig. 1 \& 2.

To train the neural networks, supported nanoparticles are constructed with the Atomic Simulation Environment (ASE) \cite{HjorthLarsen2017TheAtoms} and HR-TEM images are simulated using abTEM \cite{Madsen2020Simulation}. Refer to Fig. \ref{fig:Au_visual} for examples. For this work, Au face centred cubic structures of varying sizes are generated so that the [110] crystal direction is aligned with the optical axis, with a slight random tilt off zone axis of up to 3$^\circ$. At a random layer from the centre of the nanoparticle the [111] direction is truncated, effectively slicing a (111) plane. The exposed face of the FCC structure is attached to a (111) plane of a CeO$_\text{2}$ slab. These systems are randomly rotated about the optical axis, and contain varying sizes of nanoparticles. This replicates realistic interfaces between Au and CeO$_\text{2}$ identified by Liu \emph{et al.} \cite{Liu2019}. The image simulation then applies the parameters shown in Table \ref{tab:parameters}. This is a relatively cheap operation, since the expensive part of the image simulation is generating the exitwave by computing the multislice algorithm \cite{Kirkland2020AdvancedMicroscopy}. Applying imaging imperfections such as the contrast transfer function (CTF) and the modulation transfer function (MTF) can be done multiple times on the same exitwave, which generates multiple images of the same nanoparticle with varying imaging conditions. These images are pre-generated and stored to allow for a controlled comparison between networks on the exact same dataset. Training epochs cycle through the different sets of images of the same atomic system. Each set of images is referred to as an image epoch. With 300 training epochs and 10 image epochs, each image is reused 30 times in training.

Mask labels are binary images separating the Au nano-particle from the surrounding vacuum and the CeO$_\text{2}$ support. This is generated by computing the convex hull of the atomic coordinates. The neural networks are trained to map a single HR-TEM image to a binary mask label image, which will provide the pixels containing the nanoparticle separate from the pixels containing substrate and vacuum. An example of this is shown in the SOI Fig. 3. Once the network is trained it can be applied to an image, either simulated or experimental, and will return a probability map. Each pixel will be classified to belong to either the nanoparticle or the background (vacuum or substrate) class at some probability. This is referred to as the network inference or prediction and a threshold is applied at 0.9 to generate a binary predicted mask of the pixels that the network classify with at least 90\% confidence.

We note that the networks trained in this work are not trained to be applied in a generic sense. The networks are trained for Au nanoparticles specifically. We have seen that the networks will show weak predictions for nanoparticles with a similar lattice constant. This is because the network learns to separate patterns, which in this context is the atomic lattice where the spacing is defined by the lattice constant. Generic networks will likely show weaker performance, due to the increased complexity of varying lattice constants.

\begin{table}[!htb]
  \centering
  \begin{tabular}{lccl}
    \hline\hline
    Parameter & L & U & Unit \\
    \hline
    Acceleration voltage & \multicolumn{2}{c}{300} & keV \\
    Defocus ($\Delta f$) & -200 & 200 & \AA\\
    Spherical aberration  ($C_s$) & 0 & 12.45 & $\mu$m \\
    Focal spread & 5 & 20 & \AA \\
    Blur & 0.1 & 0.8 & \AA \\
    Frame dose & $10^{2}$ & $10^{6}$ & $e^-/\text{\AA}^{2}$ \\
    Resolution & 0.07 & 0.08 & \AA/pixel\\
    \hline\hline
  \end{tabular}
  \caption{Microscope parameters.  For each image series, a set of
    microscope parameters are drawn within the limits given here,
    except the acceleration voltage which is kept constant. L and U denote the lower and upper limits, respectively.
    All distributions are uniform, except for the electron dose which is
    exponential.}
  \label{tab:parameters}
\end{table}

Experimental images of a CeO$_2$ supported Au nanoparticle were acquired on an image corrected FEI Titan 80-300 ECELL at 300 keV, using a Gatan OneView detector. The Au nanoparticle is imaged initially at a frame dose of 10 $e^-/\text{\AA}^{2}$, where noise dominates the images. The doserate is continuously increased to above 1000 $e^-/\text{\AA}^{2}$, where atomic columns are relatively visible. The Au nanoparticle is situated in vacuum at 200$^\circ C$, so it is expected that the nanoparticle is relatively inert. We refer to work by Lomholdt \emph{et al.} \cite{LomholdtToPublished} for experimental details and details of the SNR at varying frame doses. 

Our approach here is to utilise the segmentation of the final frame of this continuously increasing doserate series as a pseudo-ground truth to gauge the performance of the network at lower frame doses. Due to drift in the images of the nanoparticle, pixel-wise scores such as the F1-Score will not be used, instead we will measure the area of segmentation, where the ground truth area will be the target.

For scintillating material based detectors, such as the Gatan OneView, the approximation of pure Poissonian noise in HR-TEM images breaks down, since the spectral profile of the noise is altered by the modulation transfer function (MTF) \cite{McMullan2014ComparisonMicroscopy, Faruqi2018DirectMicroscopy}. This function is the Fourier transform of the point spread function, which is an intrinsic property of the scintillating material \cite{Vulovic2010AMicroscopy}. Here we study the parametric form from Lee \emph{et al.} \cite{Lee2014ElectronImages}, shown in Eq. \ref{eq:mtf}.
\begin{equation}
    MTF(\tilde{q}) = (1-C) \cdot \frac{1}{1+(\frac{\tilde{q}}{c_0})^{c_3}} + C
    \label{eq:mtf}
\end{equation}
where the spatial frequencies are normalised by the Nyquist frequency, which is related to the sampling, $s$ of the detector by $\tilde{q} = q/q_N = 2 \cdot q \cdot s$. The limits of the function are $1$ for $\tilde{q} \rightarrow 0$ due to a normalisation and $C$ for $q \rightarrow \infty$. The function is fitted following the noise method described in Ref. \cite{Vulovic2010AMicroscopy}.

\section{Results}
\label{sec:results-discussion}

The following section will present results presenting how to achieve reliable and robust low dose segmentations of HR-TEM images. This will be divided into firstly a comparison of two neural network architectures and their ability to differentiate signal from noise. The neural networks will be trained on simulated data with different electron dose ranges to gauge their abilities to learn from training datasets of varying difficulty. After this a detailed analysis on how to tune performance by optimising the MTF will be presented.

\subsection{Dataset Dose Limits and Neural Network Comparison}
An obvious first step in approaching segmentation of low SNR data is to investigate the performance of various architectures. Here we use the experimental image series to benchmark the U-net against the MSD-net and identify the best neural network for low SNR HR-TEM segmentation. We test the neural networks abilities to train on two different ranges of frame dose presented in Table \ref{tab:doseparameters}. The low frame dose range covers the range of the experimental data series in this work, where-as the high frame dose range feeds the network much clearer simulated images, making it an easier dataset to learn. It is of interest to determine whether the neural networks ability to differentiate signal from noise benefits more from seeing primarily noisy images or clearer images of the atomic structure.

\begin{table}[!htb]
  \centering
  \begin{tabular}{lccl}
    \hline\hline
    Parameter & L & U & Unit \\
    \hline
    High frame dose range & $10^{2}$ & $10^{6}$ & $e^-/\text{\AA}^{2}$ \\
    Low frame dose range & $10^{1}$ & $10^{4}$ & $e^-/\text{\AA}^{2}$ \\
    \hline\hline
  \end{tabular}
  \caption{Frame dose ranges for simulated images.}
  \label{tab:doseparameters}
\end{table}

Comparisons will be made by plotting the segmented area versus frame dose to identify when each network begins to detect the nanoparticle and when the segmented area converges \emph{i.e.} the network identifies the entire nanoparticle. The final frame segmentation should be understood as a a pseudo-ground truth. The aim is to segment a similar area as in the ground truth at as low frame dose as possible. The area may not be exact due to possible morphological changes in the nanoparticle. The morphological changes and slight drift disallows the use of, for example, an F1-Score against the given frame and the ground truth for the experimental data series, since the overlap of the segmentations will not be sensible at a pixel-wise level. The ideal case of these plots will be a step function, meaning the entire nanoparticle is immediately identified, at some minimal frame dose.

\begin{figure*}[htp]
    \centering
    \includegraphics[width=\linewidth]{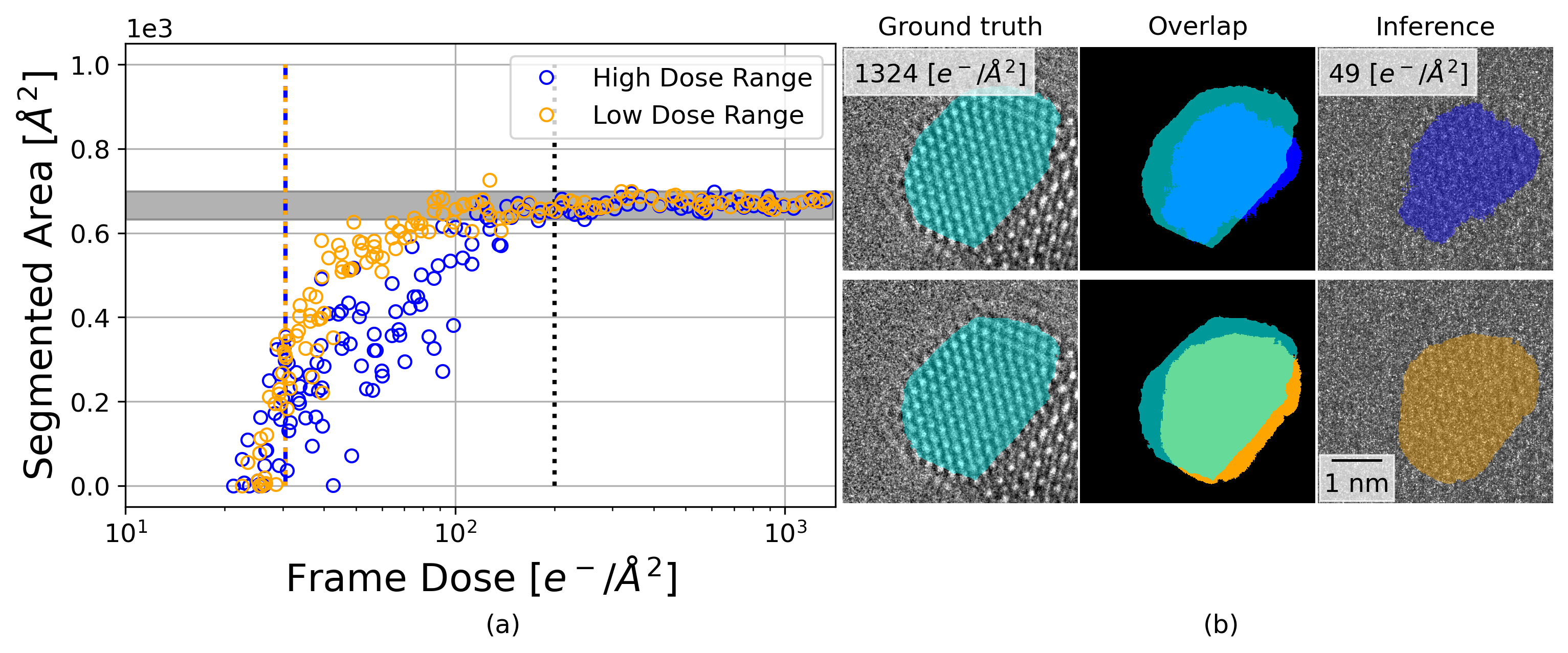}
    \caption{Comparison of the MSD-net trained with simulated images within the high frame dose range to simulated images within the low frame dose range (see Table \ref{tab:doseparameters}). (a) Segmented area on the experimental HR-TEM images by each model as a function of frame dose. Black dashed line represents the convergence of the area of segmentation, symbolising where the network segmented the entire nanoparticle. The minimum and maximum area beyond this point form the shaded grey bar as a visual aid for the target area of segmentation. Colour coded dashed lines for each model is shown representing the frame dose at which the model achieves 50\% segmentation of the nanoparticle. (b) Colour coded examples of the segmentation at 49 $e^-/\text{\AA}^{2}$ overlapped with the segmentation of the final frame (ground truth coloured in cyan). Top: The high dose range trained MSD-net segmentation and Bottom: The low dose range trained MSD-net segmentation. This highlights the ability to achieve low dose segmentation.}
    \label{fig:doserangecomparison_MSDnet}
\end{figure*}

The MSD-net presents an ability to generalise outside of the training data range, whereas the U-net only performs within the training data range. Training the MSD-net on the low frame dose range increases the visibility of the nanoparticle between 40-100 $e^-/\text{\AA}^{2}$ by $\sim$50\%, as seen in Fig. \ref{fig:doserangecomparison_MSDnet}(a), however it is also a noteworthy feature that the MSD-net is able to segment significant regions below the lower limit of the high frame dose range when trained on the high frame dose range. This proves a superior ability to separate noise to signal and generalise beyond the limits of the training set, which means the MSD-net is a better candidate for when data is limited. Fig. \ref{fig:doserangecomparison_MSDnet}(b) visualises the improvement in low dose segmentation by overlapping the ground truth segmentation (cyan coloured mask), with the segmentation from the two networks trained on each dose range (colour coded mask). The MSD-net trained on both dose ranges show a lower limit at 20-30 $e^-/\text{\AA}^{2}$.

\begin{figure*}[htp]
    \centering
    \includegraphics[width=\linewidth]{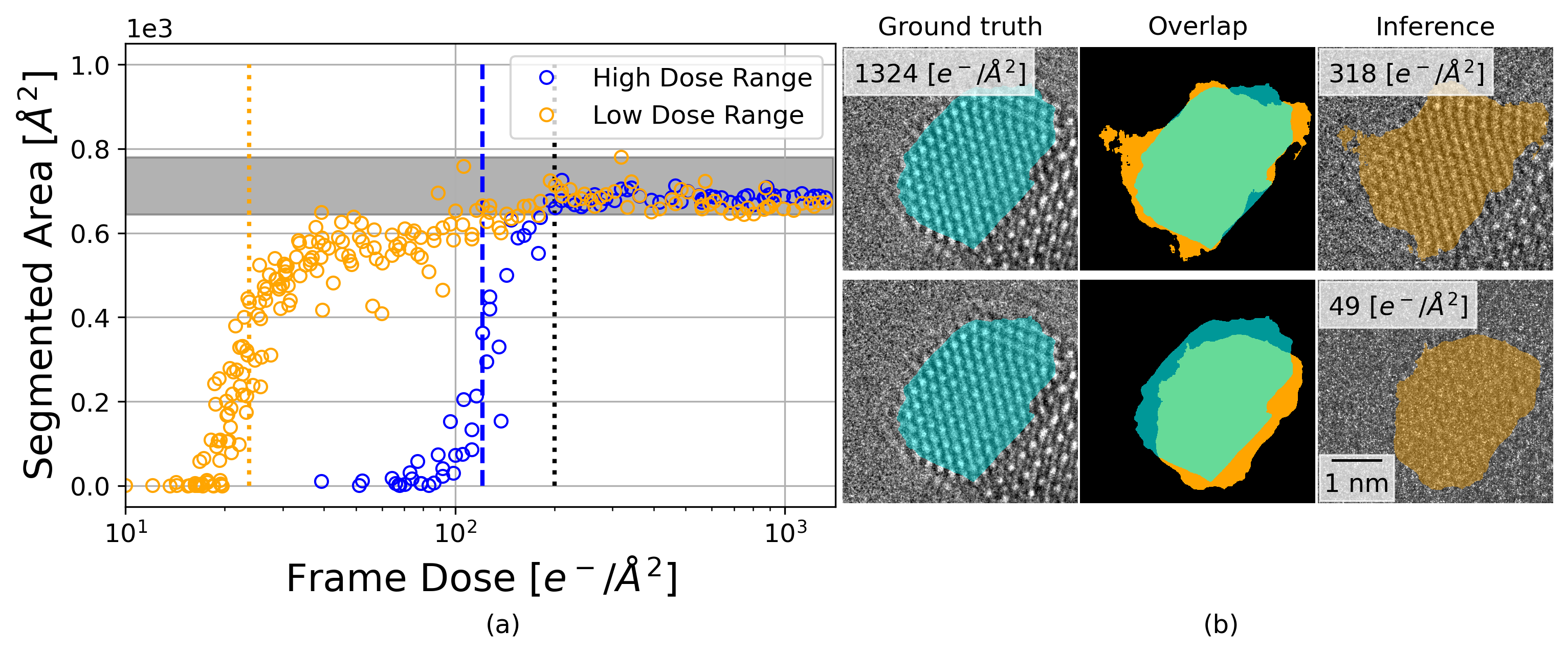}
    \caption{Comparison of the U-net trained with simulated images within the high frame dose range to simulated images within the low frame dose range (see Table \ref{tab:doseparameters}). (a) Segmented area on the experimental HR-TEM images by each model as a function of frame dose. Black dashed line represents the convergence of the area of segmentation, symbolising where the network segmented the entire nanoparticle. The minimum and maximum area beyond this point from the shaded grey bar as a visual aid for the target area of segmentation. Colour coded dashed lines for each model is shown representing the frame dose at which the model achieves 50\% segmentation of the nanoparticle. (b) Top: maximum segmentation overlapped with the segmentation of the final frame (ground truth). This highlights any over-segmented areas, due to difficulties in defining the boundaries of the nanoparticle. Bottom: Segmentation at 49 $e^-/\text{\AA}^{2}$ overlapped with the segmentation of the final frame (ground truth). This highlights the ability to achieve low dose segmentation.}
    \label{fig:doserangecomparison_Unet}
\end{figure*}

Predictions from the U-net are not as robust as the MSD-net. Fig. \ref{fig:doserangecomparison_Unet}(b) shows that the U-net trained on the low frame dose range is not entirely reliable due to the difficulties in defining the edges of the nanoparticle. Comparing the low dose trained MSD-net segmentation at 49 $e^-/\text{\AA}^{2}$ in Fig. \ref{fig:doserangecomparison_MSDnet}(b) to the low dose trained U-net at 49 $e^-/\text{\AA}^{2}$ in Fig. \ref{fig:doserangecomparison_Unet}(b), the segmented area is similar however the edges are more well defined in the segmentation from the MSD-net.

\begin{figure}[!htb]
    \centering
    \includegraphics[width=0.95\linewidth]{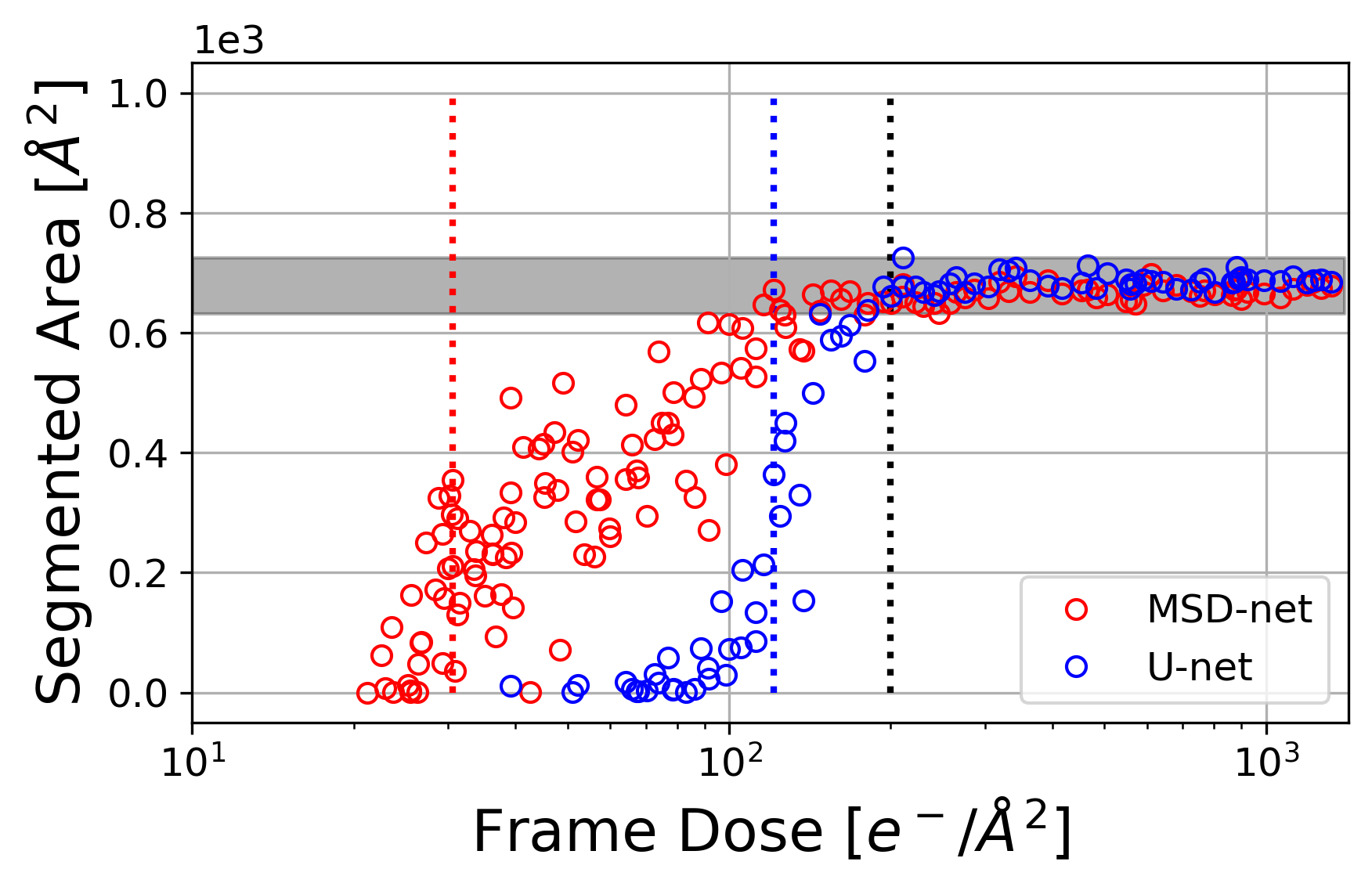}
    \caption{Comparison of the MSD-net's and U-net's ability to learn to distinguish between signal and noise. The plot shows the segmented area on the experimental HR-TEM images by each model as a function of frame dose. Black dashed line represents the convergence of the area of segmentation, symbolising where the network segmented the entire nanoparticle. The minimum and maximum area beyond this point from the shaded grey bar as a visual aid for the target area of segmentation. Colour coded dashed lines for each model is shown representing the frame dose at which the model achieves 50\% segmentation of the nanoparticle.}
    \label{fig:UvsMSD}
\end{figure}

The low dose trained U-net achieves some visibility a few $e^-/\text{\AA}^{2}$ below the MSD-net, as seen in Fig. \ref{fig:doserangecomparison_Unet}, however the over segmentation at higher dose highlights that this network is possibly being triggered by noise as well. In SOI Fig. 4 the F1-Score as a function of training epochs with simulated data highlight that the low dose range dataset becomes a harder problem for both neural networks to learn and generalise. Both networks show greater signs of overfitting. The MSD-net seems to handle more difficult problems better than the U-net. We have tested both networks on the full dose range of $10^1$-$10^6$ $e^-/\text{\AA}^{2}$ and find that the over segmentation artefacts from the low frame dose range remain. This furthers the argument that the networks benefit more from the clearer images of the atomic structures. As a sanity check, we show the effect of various training dataset sizes in SOI Fig. 5. Finally Fig. \ref{fig:UvsMSD} compares the two networks trained on the high frame dose range data, showcasing the performance gain in the low frame dose regime with the MSD-net segmenting 50\% of the nanoparticle at a $\sim$70\% lower frame dose, and both show a reliable convergence at 200 $e^-/\text{\AA}^{2}$.

\subsection{Frame dose limit}

\begin{figure}[!htb]
  \centering
  \includegraphics[width=0.98\linewidth]{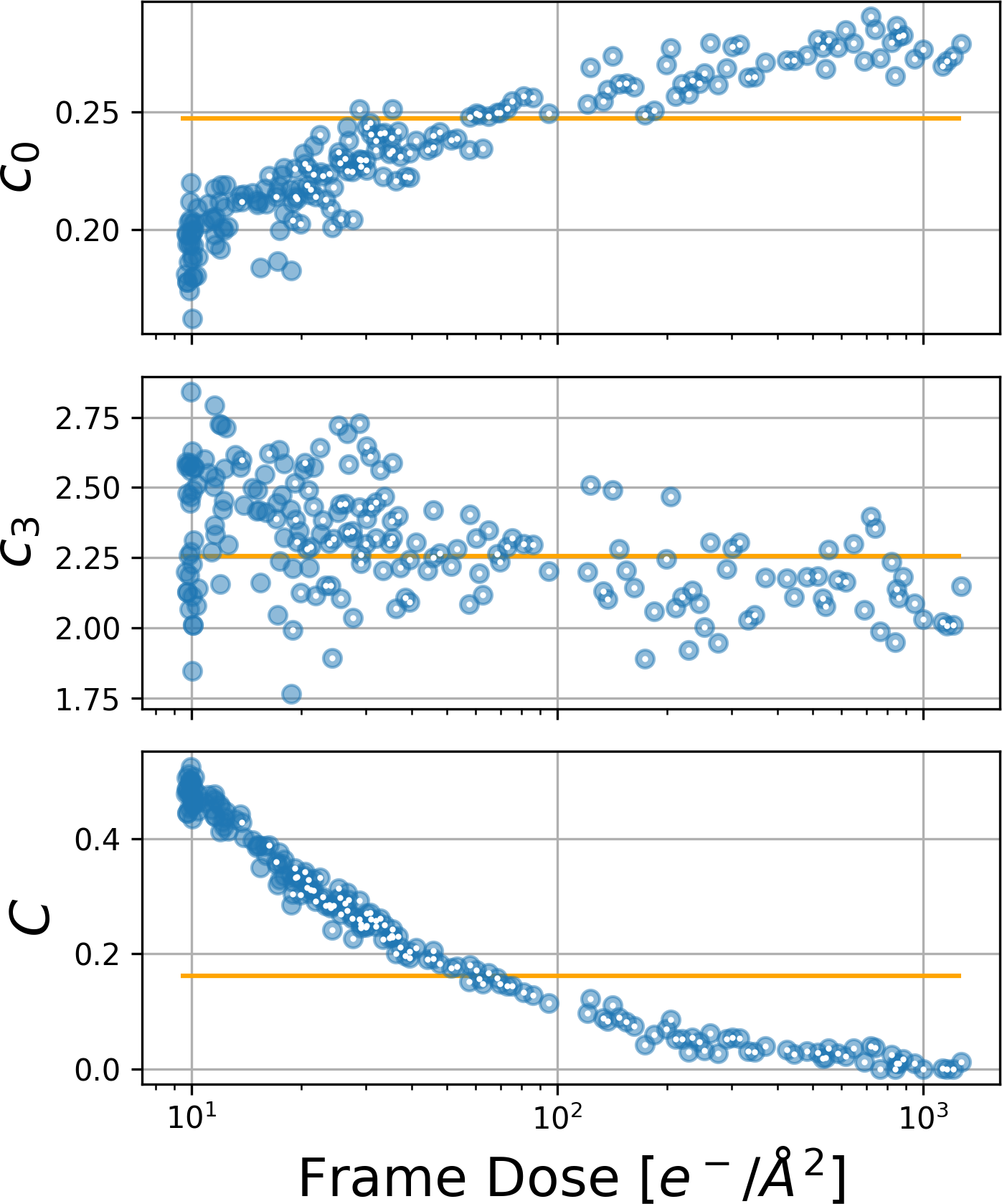}
  \caption{The fitted parameters of Eq. \ref{eq:mtf} for each frame in the experimental data series. Fits with $R^2 \geq 0.98$ are marked by a white dot. The orange line represents the mean value.}
  \label{fig:mtfvsdose}
\end{figure}

Low dose segmentation performance relies on a proper modelling of the noise in the simulated data. The following will described how noise characteristics are extracted from experimental data and modelled in simulated data. Eq. \ref{eq:mtf} was fitted to the azimuthally averaged centred Fourier transform of the vacuum region in the images, as done in \cite{Vulovic2010AMicroscopy, Lee2014ElectronImages}. See SOI Fig. 6 \& 7 for the fitted MTF of the first frame and last frame of the experimental image series, respectively. Fig. \ref{fig:mtfvsdose} presents a distribution of fitted parameters dependent on the frame dose of the experimental images. The parameter points with white dots are the fits with an $R^2 \geq 0.98$.

It is vital to understand the role of each parameter in Eq. \ref{eq:mtf}, to interpret the distributions in Fig. \ref{fig:mtfvsdose}. The $c_0$ parameter represents the $\tilde{q}$ value of the half-maximum, which increases with frame dose and converges around 200 $e^-/\text{\AA}^{2}$. This means that at frame doses above 200 $e^-/\text{\AA}^{2}$ spatial frequencies up to $\sim 0.27 \cdot q_N$ are maintained at at least half maximum, but for lower doses the function is narrower.  As a result spatial frequencies above $c_0\cdot q_N$ are effectively filtered out at higher frame doses.  The curvature of the function is consistently close to a Lorentzian form, as revealed by the $c_3$ parameter (higher values approach a low pass step function).  The $C$ parameter reveals the change in the tail of the function \emph{i.e.} the $q \rightarrow \infty$ limit.  $C$ is non-zero at lower frame dose and approaches 0 at higher doses.  We interpret this as a fraction of the noise that is not subject to the point spread function of the scintillator, but is generated later in the detection process.  We label this part of the noise ``readout noise'' although it may come from more than one source. 

All parameters show most variation below 200 $e^-/\text{\AA}^{2}$. The variation below this limit is likely due to the transition between readout noise and shot noise as the dominating noise source \cite{DeRuijter1992MethodsDetection}.

\begin{figure}[!htb]
  \centering
  \includegraphics[width=0.95\linewidth]{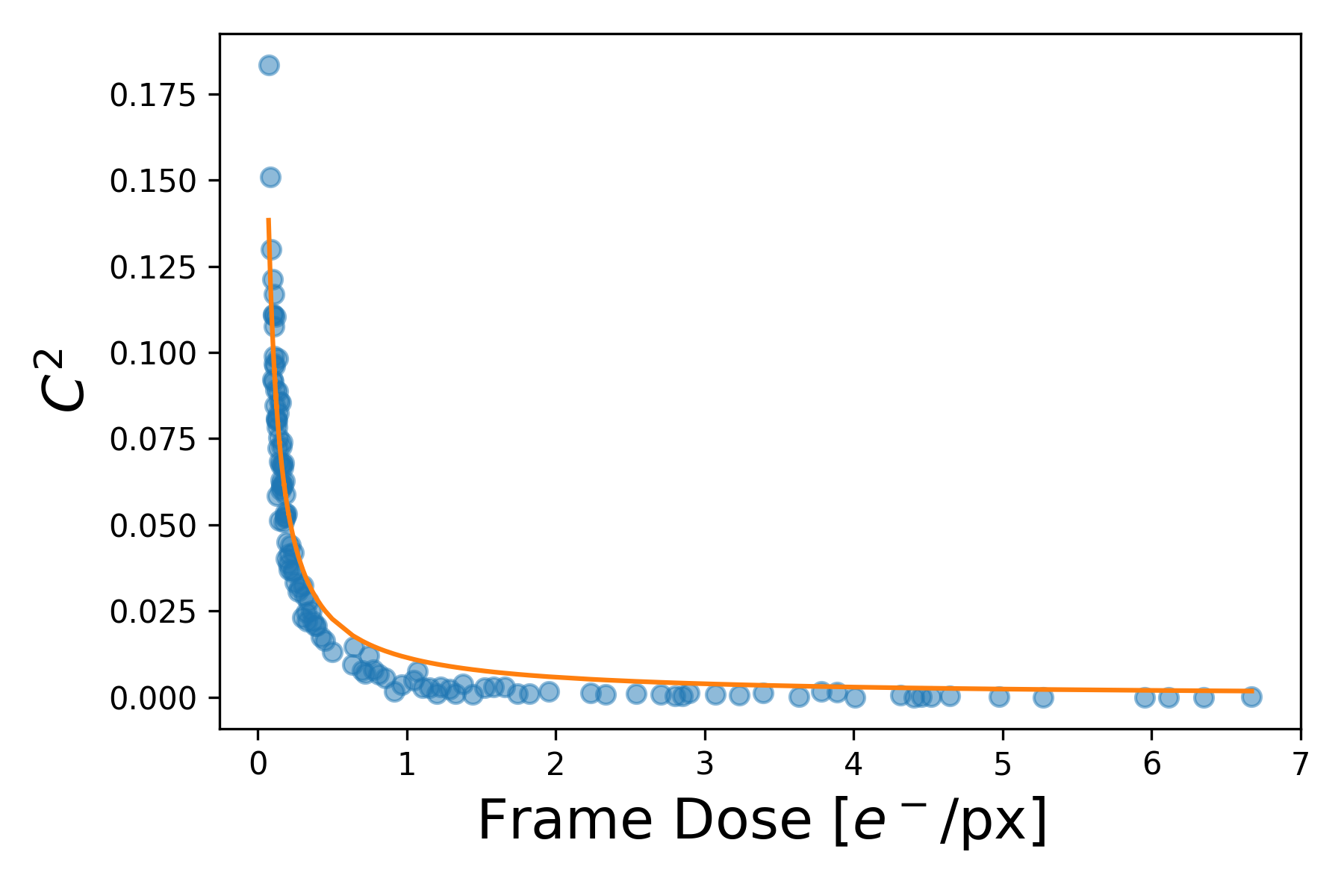}
  \caption{The correlation of the $C$ parameter of Eq. \ref{eq:mtf} to the frame dose as electrons per pixel. The fraction of readout noise, $N_0$ is extracted by fitting Eq. \ref{eq:N0Fit}. The orange lines shows the fitted curve with $R^2 = 0.94$ and $N_0=0.01$.}
  \label{fig:pcorrelation}
\end{figure}

The readout noise appears after the scintillating material and is therefore not affected by the MTF. The contributions of the readout noise and shot noise are modelled separately in our simulations and in order to extract the fractional contribution of each noise source we look at how $C$ dependends on the frame dose (in electrons per pixel). This value represents the readout noise, a minimal noise always present in the data. At higher frame dose readout noise is hidden by the shot noise, but at lower dose the readout noise dominates. The shot noise is modelled as a Poisson distribution, $P(N_D)$, where $N_D$ is the frame dose in electrons per pixel.  For lack of better models, the read out noise is also modelled as a Poisson distribution, $P(N_0)$.  $N_0$ can be therefore be extracted by fitting the $N_D$ dependency of $C$ with Eq. \ref{eq:N0Fit}, see Fig. \ref{fig:pcorrelation}. 
\begin{equation}\label{eq:N0Fit}
    C^2 = \frac{N_0}{N_0 + N_D} .
\end{equation}
We thus model the total noise as a sum of two Poisson distributions, one representing the shot noise spread by the MTF, the other represents the readout noise:
\begin{equation}
    P_{total} = MTF[P(N_D)] + P(N_0) .
\end{equation}

The approach we have taken to model the variations of $c_0$ and $c_3$ is to randomise the parameters within a given range as done by Madsen \emph{et al}. \cite{Madsen2018AImages}, which alters the spectral profile of the shot noise. The extracted $N_0$ from \ref{fig:pcorrelation} is varied at $\pm$50\%, \emph{i.e.} $N_0\in[0.005, 0.015]$ when applied to simulated images. We acknowledge that we are not accounting for all noise contributions and that noise is by nature difficult to reproduce. Varying $N_0$ accounts for deviations from our simplified noise model. In SOI Fig. 8, we show that with the fitted $N_0$ applied as a separate Poissonian noise source we are able to replicate the $C$ parameter dependency on the frame dose from MTFs fitted to a simulated image series of vacuum at increasing frame dose. The parameters $c_0$ and $c_3$ show a larger spread in the simulation compared to the experimentally fitted values. We have neglected the modelling of these parameters and they are simply ranged within a uniform distribution, as presented in SOI Fig. 8. The effect of altering this range is explored in Table \ref{tab:MTFparameters}.

The MTF is an intrinsic property of the specific detector being used. It is of interest to identify the importance of modelling the exact MTF or if a range of varying MTF profiles can be applied. To study the effect of the MTF spread on neural network training, the ranges of $c_0$ and $c_3$ can be varied, while maintaining a constant range of readout noise defined by $N_0$. The $c_0$ parameters controls the spatial frequency dependent half-maximum of the MTF. Spatial frequencies beyond $c_0$ are heavily damped. The $c_3$ parameter alters the curvature of the MTF profile, which affects the spatial frequencies near the half-maximum. Table \ref{tab:MTFparameters} summarises 5 different MTF parameter ranges applied to 5 simulated datasets of identical atomic systems. All apply the microscope parameters from Table \ref{tab:parameters}. An identical MSD-net is trained on each dataset. The first model, MTF 1, will take the parameters from Ref. \cite{Madsen2018AImages}. Motivations behind the other MTF models will be explained as the results are discussed.

\begin{table}[!htb]
  \centering
  \begin{tabular}{c|cc|cc|cc}
    \hline\hline
    \multirow{2}{*}{MTF} & \multicolumn{2}{c|}{$c_0$} & \multicolumn{2}{c|}{$c_3$} & \multicolumn{2}{c}{$N_0$} \\
    & L & U & L & U & L & U \\
    \hline
    $1$ & 0.4 & 0.6 & 2.0 & 3.0 & 0.005 & 0.015 \\
    $2$ & 0.4 & 0.6 & 1.0 & 5.0 & 0.005 & 0.015 \\
    $3$ & 0.2 & 0.3 & 1.0 & 5.0 & 0.005 & 0.015 \\
    $4$ & 0.25 & 0.6 & 2.0 & 5.0 & 0.005 & 0.015 \\
    $5$ & 0.6 & 0.8 & 2.0 & 5.0 & 0.005 & 0.015 \\
    \hline\hline
  \end{tabular}
  \caption{The various MTF models studied in the work. 1-5 represents different ranges of parameters in Eq. \ref{eq:mtf}.}
  \label{tab:MTFparameters}
\end{table}

\begin{figure*}[htp]
    \centering
    \includegraphics[width=\linewidth]{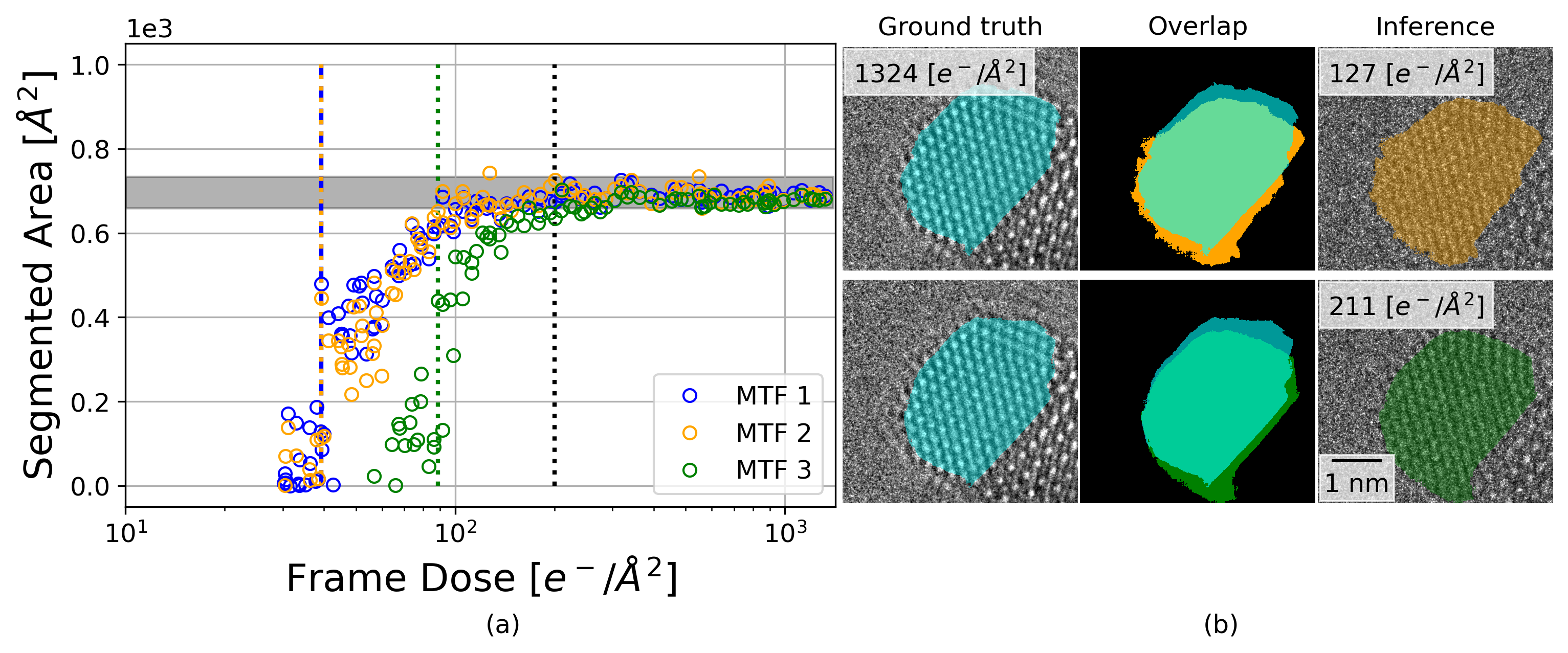}
    \caption{Comparison of the first 3 MTF models presented in Table \ref{tab:MTFparameters}. (a) Segmented area on the experimental HR-TEM images by each model as a function of frame dose. Black dashed line represents the convergence of the area of segmentation, symbolising where the network segmented the entire nanoparticle. The minimum and maximum area beyond this point from the shaded grey bar as a visual aid for the target area of segmentation. Colour coded dashed lines for each model is shown representing the frame dose at which the model achieves 50\% segmentation of the nanoparticle. MTF 2 shows the best performance (most nanoparticle visibility at the lowest frame dose). (b) Colour coded examples of the maximum segmentation overlapped with the segmentation of the final frame (ground truth). This highlights any over-segmented areas, due to difficulties in defining the boundaries of the nanoparticle.}
    \label{fig:mtfcomparison_1}
\end{figure*}

Higher values of $c_0$ in MTF 1 and MTF 2 assist the network in low frame dose segmentation. Fig. \ref{fig:mtfcomparison_1}(a) presents the visibility of the entire nanoparticle as a function of frame dose from 3 MSD-net models trained on MTF models 1-3 individually. MTF 1 and MTF 2 show very similar performance, which highlights that it is not important that the form of the MTF is exactly Lorentzian. This was determined by ranging $c_3$ over a larger range in MTF 2. Following this result, MTF 3 samples a range of $c_0$ more suited to the fitted ranges in Fig. \ref{fig:mtfvsdose}. The result of this range were detrimental on the low dose performance, delaying the visibility of the nanoparticle by $\sim$25\% in electron dose. 

Applying the range of $c_0$ within the experimental fit by MTF 3 provide more refined segmentations at higher frame dose. The variations in the segmented area at higher doses for the MTF 3 trained model are smaller than that of the MTF 1 and 2. The variations of MTF 1 and MTF 2 are shown by the shaded grey region. These variations can be due to difficulties in defining the boundaries between the nanoparticle and substrate/vacuum in the images. Fig. \ref{fig:mtfcomparison_1}(b) shows the maximum segmented area of MTF 2 (top) and MTF 3 (bottom). Here it is seen that the MTF 2 trained model has minor difficulties in defining the border of the nanoparticles.

We speculate that preserving spatial frequencies up to around half-Nyquist is preferential for the networks to learn to differentiate signal from noise. In practice this means having a $c_0$ that ranges up to or slightly above $0.5\cdot q_N$. Values of $c_0$ within the fitted range from experimental data is however also necessary for refining boundaries of the segmented areas.

\begin{figure*}[htp]
    \centering
    \includegraphics[width=\linewidth]{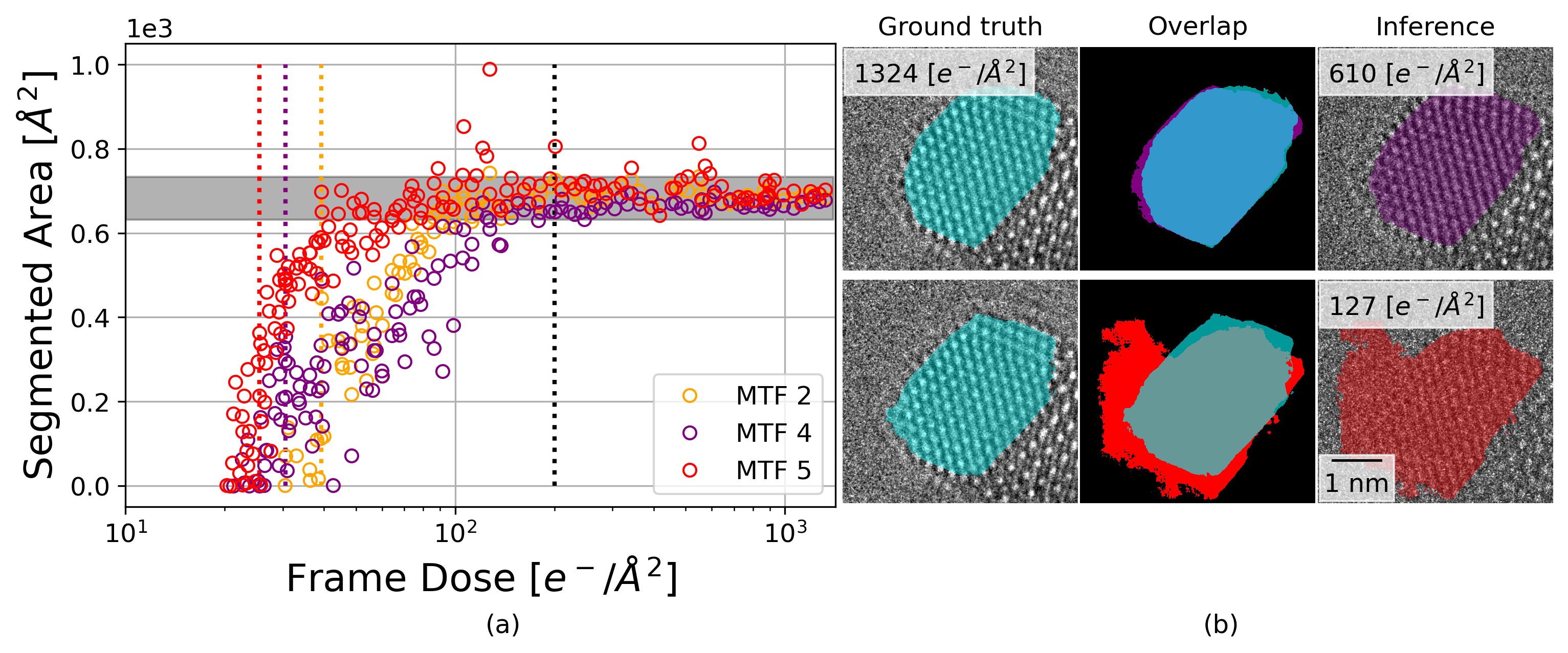}
    \caption{Comparison of the last 3 MTF models presented in Table \ref{tab:MTFparameters}. (a) Segmented area on the experimental HR-TEM images by each model as a function of frame dose. Black dashed line represents the convergence of the area of segmentation, symbolising where the network segmented the entire nanoparticle. The minimum and maximum area beyond this point from the shaded grey bar as a visual aid for the target area of segmentation. Colour coded dashed lines for each model is shown representing the frame dose at which the model achieves 50\% segmentation of the nanoparticle. MTF 4 shows the best performance (most nanoparticle visibility at the lowest frame dose and a tight convergence of the segmented area). (b) Colour coded examples of the maximum segmentation overlapped with the segmentation of the final frame (ground truth). This highlights any over-segmented areas, due to difficulties in defining the boundaries of the nanoparticle.}
    \label{fig:mtfcomparison_2}
\end{figure*}

MTF 4 ranges between the lower limit of the fitted $c_0$ from MTF 3 and the upper limits of $c_0$ from MTF 1 and 2, and achieves the low dose performance of MTF 1 and 2 and the refined boundaries of MTF 3 at higher dose. MTF 5 samples larger $c_0$, beyond half-Nyquist, with the same range as MTF 1 and 2. Fig. \ref{fig:mtfcomparison_2}(a) presents the performance of MTF models 2,4, and 5. MTF 5 shows much more visibility at low dose frames, however much more sporadic variations in the higher frame dose regime. Fig. \ref{fig:mtfcomparison_2}(b) shows the maximum segmented area of MTF 5, which highlights its weakness in identifying the borders of the nanoparticle; The segmentation bleeds into the surrounding vacuum. This renders the low dose segmentations of MTF 5 unreliable, as it seems it is being triggered by noise in the vacuum. MTF 5 has such a high $c_0$ range that it approaches a noise profile appropriate for direct electron counting detectors such as the Gatan K2/3 camera \cite{Faruqi2018DirectMicroscopy}. The maximum segmentation of MTF 4 in contrast shows very sharp separations between the nanoparticle and its surroundings. This proves that ranging $c_0$ such that it covers the range fitted from experimental data, but also such that it retains spatial frequencies up to half-Nyquist is ideal for optimal segmentation performance across the entire frame dose range.

We note that in all cases the area of segmentation converges at approximately 200 $e^-/\text{\AA}^{2}$, depicted by the dashed black line. This is the first reported lower limit of frame dose for a reliable full segmentation.

The overlap in the middle of Figs. \ref{fig:mtfcomparison_1}(b) and \ref{fig:mtfcomparison_2}(b) shows the morphological changes in the nanoparticle but also highlight that both are segmentations of the same nanoparticle. Here we also highlight that the segmentation is sensible and we show that this is at a level where a human interpreter would have difficulties being certain of the presence of the entire nanoparticle.

The segmentations below 200 $e^-/\text{\AA}^{2}$ are not full segmentations and cannot be used to for example measure the area, but can still be used for object detection purposes and regional Fourier transform extraction. For these purposes 50\% of segmentation would be a sensible minimum, represented by the colour coded dashed lines in Figs. \ref{fig:doserangecomparison_MSDnet}(a), \ref{fig:doserangecomparison_Unet}(a), \ref{fig:mtfcomparison_1}(a), and \ref{fig:mtfcomparison_2}(a).

To further prove the performance of the MTF 4 model, we show that an MSD-net trained with the MTF 4 model performs better on a simulated dataset with the MTF 3 model applied, compared to an MSD-net trained with the MTF 3 model. An image series was simulated to replicate the experimental series. This simulated series contains 1000 images of a CeO$_2$ supported Au nanoparticle in the [110] zone axis positioned similarly to the experimental sample, with a blur range of [0.1, 0.2] \AA, focal spread range of [5, 6] \AA, defocus range of [48, 50] \AA, C$_s$ range of [7, 8] $\mu$m, a frame dose from [$10^1$, $10^4$] $e^-/\text{\AA}^{2}$, and the MTF 3 model. In this case we have exact ground truths and the F1-Score will be used as a metric. SOI Fig. 9a presents a histogram of the F1-Score for each image in the simulated series for both networks in mention. It is immediately apparent that the MTF 4 dataset trains a network that outperforms the MTF 3 trained network on images with the MTF 3 ranges applied. SOI Fig. 9b displays the F1-Score against frame dose, which highlights the improved performance of the MTF 4 dataset trained MSD-net is primarily on low frame dose simulated images, with a higher mean F1-Score in the range of 10-100 $e^-/\text{\AA}^{2}$.

% \FloatBarrier
\section{Conclusions}
\label{sec:conclusions}
In this work we have investigated the quantitative limit of low SNR HR-TEM image segmentation. A continuously increasing frame dose HR-TEM image series of a CeO$_2$ supported Au nanoparticle was acquired to rate the performance of neural network segmentations at low frame dose and compare the segmentation to the final (highest frame dose) frame.

The results show that the neural networks are achieving human level performance, which means it can safely be equipped for large-scale automated analysis to relieve the human operator of repetitive tasks.

The MSD-net showed promising low SNR performance compared to the industry standard U-net. The MSD-net showed an ability to generalise outside the frame dose range of the training set and provide reliable segmentations, with well defined boundaries down to 20 $e^-/\text{\AA}^{2}$. The U-net was only able to operate within the given training range, and when trained with lower dose images, presented weaknesses in defining boundaries between the nanoparticle and its surroundings. This highlights the MSD-net's superior ability to learn the differences between signal and noise.

A parametric form of the MTF was fitted to all frames in the HR-TEM series resulting in a frame dose dependent range of parameters. The noise contributions were separated into two sources, shot noise and readout noise, and the fractional contribution of each was extracted from the noise floor given by the $C$ parameter of Eq. \ref{eq:mtf}. It was shown that modelling the MTF to retain spatial frequencies up to $\sim0.5\cdot q_N$ with at least half-maximum assisted the MSD-net in detecting the nanoparticle at lower frame dose. It seems that modelling the MTF with a wide range of parameters that both consist of parameters within the fitted range and parameters that maintain spatial frequencies up to half-Nyquist allows the MSD-net to operate in both the shot noise and readout noise dominated regimes.

All results converged at 200 $e^-/\text{\AA}^{2}$. This is the first report of a frame dose limit for reliable neural network segmentations. Neural network predictions with HR-TEM images between 20-100 $e^-/\text{\AA}^{2}$ are still useful for object detection purposes.

Knowledge of these frame dose limits will provide the community with realistic expectations from deep learning models and an ability to design experiments that are optimised for their needs from deep learning powered analysis tools, whether it be object detection in live low dose imaging, or statistical data accumulating of morphological properties, amongst others.

%\section*{Acknowledgements}%% if any
%Text for this section\ldots
\FloatBarrier
\section*{Funding}%% if any
The authors acknowledge financial support from the Independent
Research Fund Denmark (DFF-FTP) through grant no.~9041-00161B.

\section*{Availability of data and materials}
The code is available on GitLab \cite{NN_HRTEM}. The trained networks, the
scripts and data used to train the network and generate all figures are available from the DTU Data repository at doi:10.11583/DTU.22836338.

\section*{Competing interests}
The authors declare that they have no competing interests.


\bibliography{references.bib}      % Bibliography file (usually '*.bib' )

\begin{thebibliography}{10}
\expandafter\ifx\csname url\endcsname\relax
  \def\url#1{\texttt{#1}}\fi
\expandafter\ifx\csname urlprefix\endcsname\relax\def\urlprefix{URL }\fi
\expandafter\ifx\csname href\endcsname\relax
  \def\href#1#2{#2} \def\path#1{#1}\fi

\bibitem{Batson2008Beameffect}
P.~E. Batson, {Motion of Gold Atoms on Carbon in the Aberration-Corrected
  STEM}, Microscopy and Microanalysis 14~(1) (2008) 89--97.
\newblock \href {https://doi.org/10.1017/s1431927608080197}
  {\path{doi:10.1017/s1431927608080197}}.

\bibitem{VanDyck:2015fg}
D.~V. Dyck, I.~Lobato, F.-R. Chen, C.~Kisielowski, {Do you believe that atoms
  stay in place when you observe them in HREM?}, Micron 68 (2015) 158--163.
\newblock \href {https://doi.org/10.1016/j.micron.2014.09.003}
  {\path{doi:10.1016/j.micron.2014.09.003}}.

\bibitem{Egerton2019RadiationTEM}
R.~F. Egerton, \href{https://doi.org/10.1016/j.micron.2019.01.005}{{Radiation
  damage to organic and inorganic specimens in the TEM}}, Micron 119~(November
  2018) (2019) 72--87.
\newblock \href {https://doi.org/10.1016/j.micron.2019.01.005}
  {\path{doi:10.1016/j.micron.2019.01.005}}.
\newline\urlprefix\url{https://doi.org/10.1016/j.micron.2019.01.005}

\bibitem{Du2015AMicrographs}
H.~Du, \href{http://dx.doi.org/10.1016/j.ultramic.2014.11.012}{{A nonlinear
  filtering algorithm for denoising HR(S)TEM micrographs}}, Ultramicroscopy 151
  (2015) 62--67.
\newblock \href {https://doi.org/10.1016/j.ultramic.2014.11.012}
  {\path{doi:10.1016/j.ultramic.2014.11.012}}.
\newline\urlprefix\url{http://dx.doi.org/10.1016/j.ultramic.2014.11.012}

\bibitem{Furnival2017DenoisingThresholding}
T.~Furnival, R.~K. Leary, P.~A. Midgley,
  \href{http://dx.doi.org/10.1016/j.ultramic.2016.05.005}{{Denoising
  time-resolved microscopy image sequences with singular value thresholding}},
  Ultramicroscopy 178 (2017) 112--124.
\newblock \href {https://doi.org/10.1016/j.ultramic.2016.05.005}
  {\path{doi:10.1016/j.ultramic.2016.05.005}}.
\newline\urlprefix\url{http://dx.doi.org/10.1016/j.ultramic.2016.05.005}

\bibitem{Lin2021:TEMImageNet}
R.~Lin, R.~Zhang, C.~Wang, X.-Q. Yang, H.~L. Xin, {TEMImageNet training library
  and AtomSegNet deep-learning models for high-precision atom segmentation,
  localization, denoising, and deblurring of atomic-resolution images},
  Scientific Reports 11~(1) (2021) 5386.
\newblock \href {https://doi.org/10.1038/s41598-021-84499-w}
  {\path{doi:10.1038/s41598-021-84499-w}}.

\bibitem{Vincent2021DevelopingSignal-to-Noise}
J.~L. Vincent, R.~Manzorro, S.~Mohan, B.~Tang, D.~Y. Sheth, E.~P. Simoncelli,
  D.~S. Matteson, C.~Fernandez-Granda, P.~A. Crozier, {Developing and
  Evaluating Deep Neural Network-Based Denoising for Nanoparticle TEM Images
  with Ultra-Low Signal-to-Noise}, Microscopy and Microanalysis 27~(6) (2021)
  1431--1447.
\newblock \href {https://doi.org/10.1017/S1431927621012678}
  {\path{doi:10.1017/S1431927621012678}}.

\bibitem{Spurgeon2021TowardsMicroscopy}
S.~R. Spurgeon, C.~Ophus, L.~Jones, A.~Petford-Long, S.~V. Kalinin, M.~J.
  Olszta, R.~E. Dunin-Borkowski, N.~Salmon, K.~Hattar, W.~C.~D. Yang,
  R.~Sharma, Y.~Du, A.~Chiaramonti, H.~Zheng, E.~C. Buck, L.~Kovarik, R.~L.
  Penn, D.~Li, X.~Zhang, M.~Murayama, M.~L. Taheri,
  \href{http://dx.doi.org/10.1038/s41563-020-00833-z}{{Towards data-driven
  next-generation transmission electron microscopy}}, Nature Materials 20~(3)
  (2021) 274--279.
\newblock \href {https://doi.org/10.1038/s41563-020-00833-z}
  {\path{doi:10.1038/s41563-020-00833-z}}.
\newline\urlprefix\url{http://dx.doi.org/10.1038/s41563-020-00833-z}

\bibitem{Treder2022ApplicationsMicroscopy}
K.~P. Treder, C.~Huang, J.~S. Kim, A.~I. Kirkland, {Applications of deep
  learning in electron microscopy}, Microscopy 71~(August 2021) (2022)
  I100--I115.
\newblock \href {https://doi.org/10.1093/jmicro/dfab043}
  {\path{doi:10.1093/jmicro/dfab043}}.

\bibitem{Groschner2021MachineData}
C.~K. Groschner, C.~Choi, M.~C. Scott, {Machine Learning Pipeline for
  Segmentation and Defect Identification from High-Resolution Transmission
  Electron Microscopy Data}, Microscopy and Microanalysis 27~(3) (2021)
  549--556.
\newblock \href {https://doi.org/10.1017/S1431927621000386}
  {\path{doi:10.1017/S1431927621000386}}.

\bibitem{Faraz2022DeepStudies}
K.~Faraz, T.~Grenier, C.~Ducottet, T.~Epicier,
  \href{https://doi.org/10.1038/s41598-022-06308-2}{{Deep learning detection of
  nanoparticles and multiple object tracking of their dynamic evolution during
  in situ ETEM studies}}, Scientific Reports 12~(1) (2022) 1--15.
\newblock \href {https://doi.org/10.1038/s41598-022-06308-2}
  {\path{doi:10.1038/s41598-022-06308-2}}.
\newline\urlprefix\url{https://doi.org/10.1038/s41598-022-06308-2}

\bibitem{Madsen2018AImages}
J.~Madsen, P.~Liu, J.~Kling, J.~B. Wagner, T.~W. Hansen, O.~Winther,
  J.~Schi{\o}tz, {A Deep Learning Approach to Identify Local Structures in
  Atomic-Resolution Transmission Electron Microscopy Images}, Advanced Theory
  and Simulations 1~(8) (2018) 1--12.
\newblock \href {https://doi.org/10.1002/adts.201800037}
  {\path{doi:10.1002/adts.201800037}}.

\bibitem{Ragone2020AtomicLearning}
M.~Ragone, V.~Yurkiv, B.~Song, A.~Ramsubramanian, R.~Shahbazian-Yassar,
  F.~Mashayek, \href{https://doi.org/10.1016/j.commatsci.2020.109722}{{Atomic
  column heights detection in metallic nanoparticles using deep convolutional
  learning}}, Computational Materials Science 180~(December 2019) (2020)
  109722.
\newblock \href {https://doi.org/10.1016/j.commatsci.2020.109722}
  {\path{doi:10.1016/j.commatsci.2020.109722}}.
\newline\urlprefix\url{https://doi.org/10.1016/j.commatsci.2020.109722}

\bibitem{Horwath2020UnderstandingImages}
J.~P. Horwath, D.~N. Zakharov, R.~M{\'{e}}gret, E.~A. Stach,
  \href{http://dx.doi.org/10.1038/s41524-020-00363-x}{{Understanding important
  features of deep learning models for segmentation of high-resolution
  transmission electron microscopy images}}, npj Computational Materials 6~(1)
  (2020) 1--9.
\newblock \href {https://doi.org/10.1038/s41524-020-00363-x}
  {\path{doi:10.1038/s41524-020-00363-x}}.
\newline\urlprefix\url{http://dx.doi.org/10.1038/s41524-020-00363-x}

\bibitem{Saaim2022InSegmentation}
K.~M. Saaim, S.~K. Afridi, M.~Nisar, S.~Islam,
  \href{https://doi.org/10.1016/j.ultramic.2021.113437}{{In search of best
  automated model: Explaining nanoparticle TEM image segmentation}},
  Ultramicroscopy 233~(December 2021) (2022) 113437.
\newblock \href {https://doi.org/10.1016/j.ultramic.2021.113437}
  {\path{doi:10.1016/j.ultramic.2021.113437}}.
\newline\urlprefix\url{https://doi.org/10.1016/j.ultramic.2021.113437}

\bibitem{Liu2019}
P.~Liu, T.~Wu, J.~Madsen, J.~Schi{\o}tz, J.~B. Wagner, T.~W. Hansen,
  {Transformations of supported gold nanoparticles observed by: In situ
  electron microscopy}, Nanoscale 11~(24) (2019) 11885--11891.
\newblock \href {https://doi.org/10.1039/c9nr02731a}
  {\path{doi:10.1039/c9nr02731a}}.

\bibitem{Ronneberger:2015gk}
O.~Ronneberger, P.~Fischer, T.~Brox, {U-Net: Convolutional Networks for
  Biomedical Image Segmentation}, Vol. 9351 of Medical Image Computing and
  Computer-Assisted Intervention – MICCAI 2015, Springer International
  Publishing, 2015, pp. 234 -- 241.
\newblock \href {https://doi.org/10.1007/978-3-319-24574-4\_28}
  {\path{doi:10.1007/978-3-319-24574-4\_28}}.

\bibitem{LethLarsen2022ReconstructingLearning}
M.~H. Leth~Larsen, F.~Dahl, L.~P. Hansen, B.~Barton, C.~Kisielowski, S.~Helveg,
  O.~Winther, T.~W. Hansen, J.~Schi{\o}tz,
  \href{https://doi.org/10.1016/j.ultramic.2022.113641}{{Reconstructing the
  exit wave of 2D materials in high-resolution transmission electron microscopy
  using machine learning}}, Ultramicroscopy 243~(November 2022) (2022) 113641.
\newblock \href {https://doi.org/10.1016/j.ultramic.2022.113641}
  {\path{doi:10.1016/j.ultramic.2022.113641}}.
\newline\urlprefix\url{https://doi.org/10.1016/j.ultramic.2022.113641}

\bibitem{Pelt2017}
D.~M. Pelt, J.~A. Sethian, {A mixed-scale dense convolutional neural network
  for image analysis}, Proceedings of the National Academy of Sciences of the
  United States of America 115~(2) (2017) 254--259.
\newblock \href {https://doi.org/10.1073/pnas.1715832114}
  {\path{doi:10.1073/pnas.1715832114}}.

\bibitem{HjorthLarsen2017TheAtoms}
A.~Hjorth~Larsen, J.~J{\O}rgen~Mortensen, J.~Blomqvist, I.~E. Castelli,
  R.~Christensen, M.~Du{\l}ak, J.~Friis, M.~N. Groves, B.~Hammer, C.~Hargus,
  E.~D. Hermes, P.~C. Jennings, P.~Bjerre~Jensen, J.~Kermode, J.~R. Kitchin,
  E.~Leonhard~Kolsbjerg, J.~Kubal, K.~Kaasbjerg, S.~Lysgaard,
  J.~Bergmann~Maronsson, T.~Maxson, T.~Olsen, L.~Pastewka, A.~Peterson,
  C.~Rostgaard, J.~Schi{\O}tz, O.~Sch{\"{u}}tt, M.~Strange, K.~S. Thygesen,
  T.~Vegge, L.~Vilhelmsen, M.~Walter, Z.~Zeng, K.~W. Jacobsen, {The atomic
  simulation environment - A Python library for working with atoms}, Journal of
  Physics Condensed Matter 29~(27) (2017).
\newblock \href {https://doi.org/10.1088/1361-648X/aa680e}
  {\path{doi:10.1088/1361-648X/aa680e}}.

\bibitem{Madsen2020Simulation}
J.~Madsen, T.~Susi, { abTEM: ab Initio Transmission Electron Microscopy Image
  Simulation }, Microscopy and Microanalysis 26~(S2) (2020) 448--450.
\newblock \href {https://doi.org/10.1017/s1431927620014701}
  {\path{doi:10.1017/s1431927620014701}}.

\bibitem{Kirkland2020AdvancedMicroscopy}
E.~J. Kirkland,
  \href{https://link.springer.com/book/10.1007/978-3-030-33260-0}{{Advanced
  Computing in Electron Microscopy}}, in: Advanced Computing in Electron
  Microscopy, 3rd Edition, Springer Cham, 2020, pp. 143--195.
\newblock \href {https://doi.org/10.1007/978-3-030-33260-0}
  {\path{doi:10.1007/978-3-030-33260-0}}.
\newline\urlprefix\url{https://link.springer.com/book/10.1007/978-3-030-33260-0}

\bibitem{LomholdtToPublished}
W.~B. Lomholdt, M.~H. Leth~Larsen, J.~Schi{\o}tz, T.~W. Hansen, {[To Be
  Published]}.

\bibitem{McMullan2014ComparisonMicroscopy}
G.~McMullan, A.~R. Faruqi, D.~Clare, R.~Henderson,
  \href{http://dx.doi.org/10.1016/j.ultramic.2014.08.002}{{Comparison of
  optimal performance at 300keV of three direct electron detectors for use in
  low dose electron microscopy}}, Ultramicroscopy 147 (2014) 156--163.
\newblock \href {https://doi.org/10.1016/j.ultramic.2014.08.002}
  {\path{doi:10.1016/j.ultramic.2014.08.002}}.
\newline\urlprefix\url{http://dx.doi.org/10.1016/j.ultramic.2014.08.002}

\bibitem{Faruqi2018DirectMicroscopy}
A.~R. Faruqi, G.~McMullan,
  \href{http://dx.doi.org/10.1016/j.nima.2017.07.037}{{Direct imaging detectors
  for electron microscopy}}, Nuclear Instruments and Methods in Physics
  Research, Section A: Accelerators, Spectrometers, Detectors and Associated
  Equipment 878~(May 2017) (2018) 180--190.
\newblock \href {https://doi.org/10.1016/j.nima.2017.07.037}
  {\path{doi:10.1016/j.nima.2017.07.037}}.
\newline\urlprefix\url{http://dx.doi.org/10.1016/j.nima.2017.07.037}

\bibitem{Vulovic2010AMicroscopy}
M.~Vulovic, B.~Rieger, L.~J. Van~Vliet, A.~J. Koster, R.~B. Ravelli, {A toolkit
  for the characterization of CCD cameras for transmission electron
  microscopy}, Acta Crystallographica Section D: Biological Crystallography
  66~(1) (2010) 97--109.
\newblock \href {https://doi.org/10.1107/S0907444909031205}
  {\path{doi:10.1107/S0907444909031205}}.

\bibitem{Lee2014ElectronImages}
Z.~Lee, H.~Rose, O.~Lehtinen, J.~Biskupek, U.~Kaiser,
  \href{http://dx.doi.org/10.1016/j.ultramic.2014.01.010}{{Electron dose
  dependence of signal-to-noise ratio, atom contrast and resolution in
  transmission electron microscope images}}, Ultramicroscopy 145 (2014) 3--12.
\newblock \href {https://doi.org/10.1016/j.ultramic.2014.01.010}
  {\path{doi:10.1016/j.ultramic.2014.01.010}}.
\newline\urlprefix\url{http://dx.doi.org/10.1016/j.ultramic.2014.01.010}

\bibitem{DeRuijter1992MethodsDetection}
W.~J. De~Ruijter, J.~K. Weiss, {Methods to measure properties of slow-scan CCD
  cameras for electron detection}, Review of Scientific Instruments 63~(10)
  (1992) 4314--4321.
\newblock \href {https://doi.org/10.1063/1.1143730}
  {\path{doi:10.1063/1.1143730}}.

\bibitem{NN_HRTEM}
\href{https://gitlab.com/matthewhelmi/NeuralNetwork_HRTEM}{{Neural Network
  Assisted HR-TEM}}.
\newline\urlprefix\url{https://gitlab.com/matthewhelmi/NeuralNetwork_HRTEM}

\end{thebibliography}

\clearpage
\appendix
\section{Supplementary Online Information}
\let\section\subsection
\setcounter{figure}{0}
\renewcommand{\thefigure}{SOI.\arabic{figure}}
%\renewcommand{\thesubsection}{SOI.\arabic{subsection}}

%\appendix
\label{sec:supplementary}
\section{Network architecture}
\label{sec:network-architecture}

\begin{figure}[H]
  \centering
  \includegraphics[width=\linewidth]{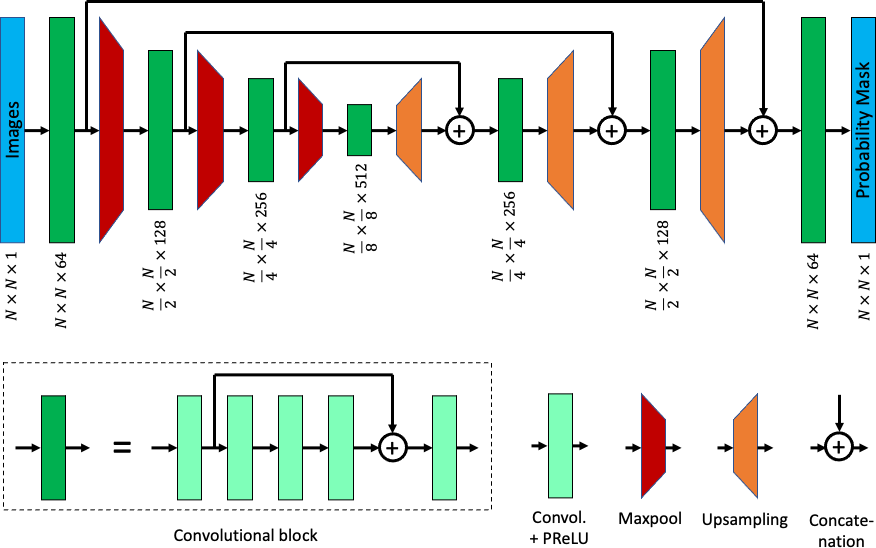}
  \caption{The architecture of the U-net.  Information flows from left to right.  The first part of the network, the ``encoding path'', consists of convolutional processing blocks alternated by downsampling layers using the MaxPool method.  The second part, the ``decoding path'', consists of convolutional blocks alternated by bilinear upsampling layers. Long skip connections ensures that the original spatial information can be maintained.  Adapted from Madsen \emph{et al.} \cite{Madsen2018AImages}}
  \label{fig:unet_arch}
\end{figure}

The U-net architecture is identical to Ref.  \cite{LethLarsen2022ReconstructingLearning}, with a difference in the initial number of channels. From hyperparameter tuning it was seen that increasing the channels minimised overfitting and provided more robust results for segmentations. The network consists of a downsampling (or ``encoding'')) path, where convolutional blocks alternate with downsampling layers, and an upsampling (or ``decoding'') path, where the convolutional blocks alternate with upsampling layers (see Fig. \ref{fig:unet_arch}).  The convolutional blocks consist of 5 convolutional layers, with a short skip connection between the output of the first layer and the input of the fifth.

The downsampling is done with conventional MaxPool operations. Each time the resolution is cut in half in a MaxPool operation, the number of feature channels in the following convolutional block is doubled to maintain the information flow in the network.  The upsampling is done using bilinear interpolation, and the following convolution block has the number of channels cut by a factor two.  After each upsampling, information from the last layer with the same spatial resolution is added from the downsampling path, this is done by concatenating the
channels.  The first layer in the convolutional blocks in
both paths will therefore have a different number of input channels from what is stated in the figure.

Each convolutional layer uses a $5 \times 5$ convolutional kernel, followed by a Parametric Leaky Rectifying Linear Unit. Hyperparameter tuning showed no significant improvement in increasing the kernel size above $5 \times 5$.

\begin{figure}[H]
  \centering
  \includegraphics[width=\linewidth]{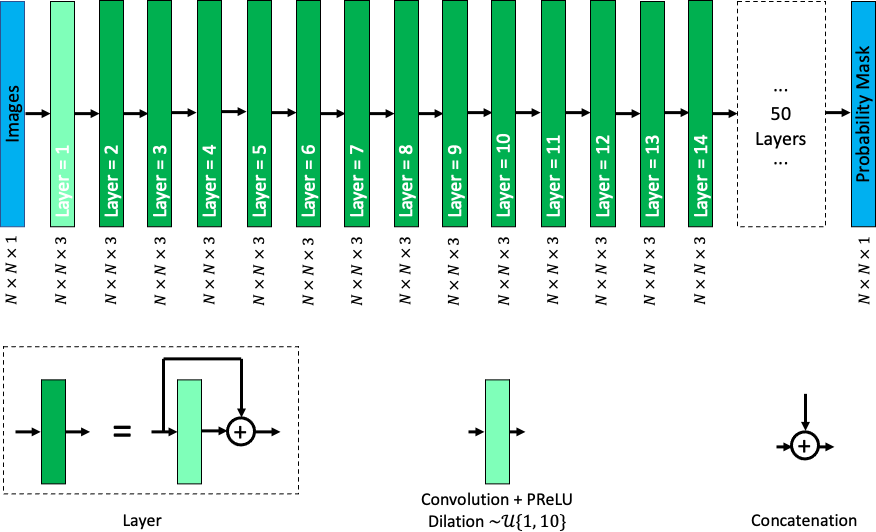}
  \caption{The architecture of the MSD-net. The convolutional kernel dilation value is sampled from a uniform distribution in the range [1,10]. Information flows from left to right.}
  \label{fig:msdnet_arch}
\end{figure}

The MSD-net architecture shown in Fig. \ref{fig:msdnet_arch}, is an adaptation of the model introduced by Pelt \emph{et al.} in Ref. \cite{Pelt2017}. The network consists of 50 layers, where each layer performs a convolution and immediately concatenates the output with the output of the previous layer. The first layer is an exception as it does not concatenate with the input image.

The reoccurring successive concatenations is what classifies this network architecture in the family of dense network architectures where all convolutional output filters are connected with eachother. This MSD-net implementation utilises a $9 \times 9$ convolutional kernel that is dilated to capture features at varying length scales. The dilation value is sampled from a uniform distribution in the range $[1,10]$, and seeding is provided to reproduce multiple identical MSD-net models. This is what ``Mixed-Scale'' refers to in the name of the architecture.

Hyperparameter tuning showed signifcant improvements when increasing the number of output channels, kernel size, and number of layers, however Tensorflow's implementation of concatenation layers consumes large amount of memory, and so parameters of this architecture are limited by available memory.
\newline

All convolutional layers apply reflection padding to maintain dimensions.

\section{Network training}
\label{sec:network-training}

\begin{figure}[H]
  \centering
  \includegraphics[width=0.98\linewidth]{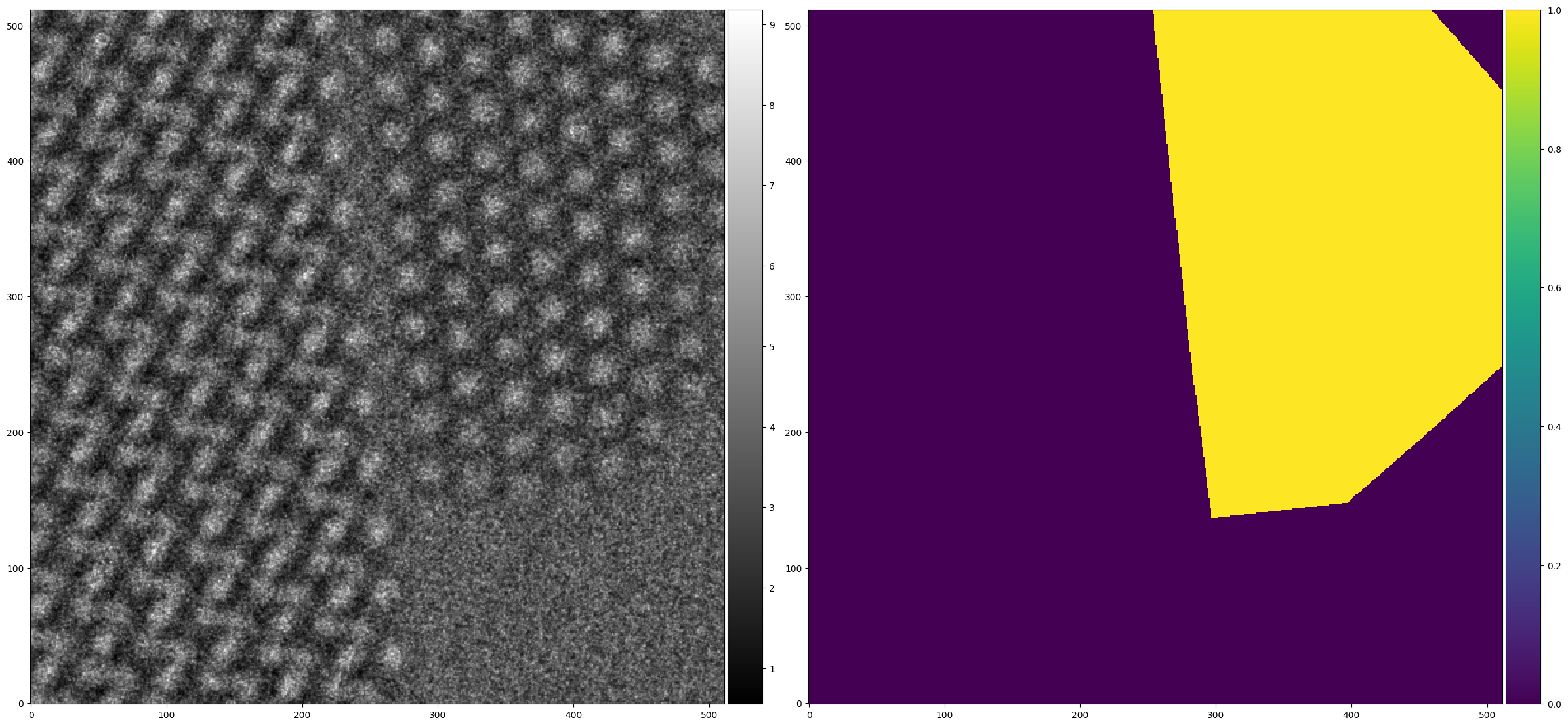}
  \caption{Example of simulated image and its binary mark label.}
  \label{fig:image_label_example}
\end{figure}

The problem definition here, as presented in Figures \ref{fig:unet_arch} and \ref{fig:msdnet_arch}, is to take a HR-TEM image and produce a probability mask of the pixels corresponding to the nanoparticle. An example of this is shown in Fig. \ref{fig:image_label_example}.

Binary masks are used so that the binary cross entropy loss (or its multi-class counterpart: the categorical cross entropy loss) can be applied for training. This is a loss function widely applied for segmentation problems and is natively implemented in Tensorflow.

\begin{figure}[H]
    \centering
    %%%%%%%%%%%%%%
    \begin{subfigure}{.49\linewidth}
        \includegraphics[width=\linewidth]{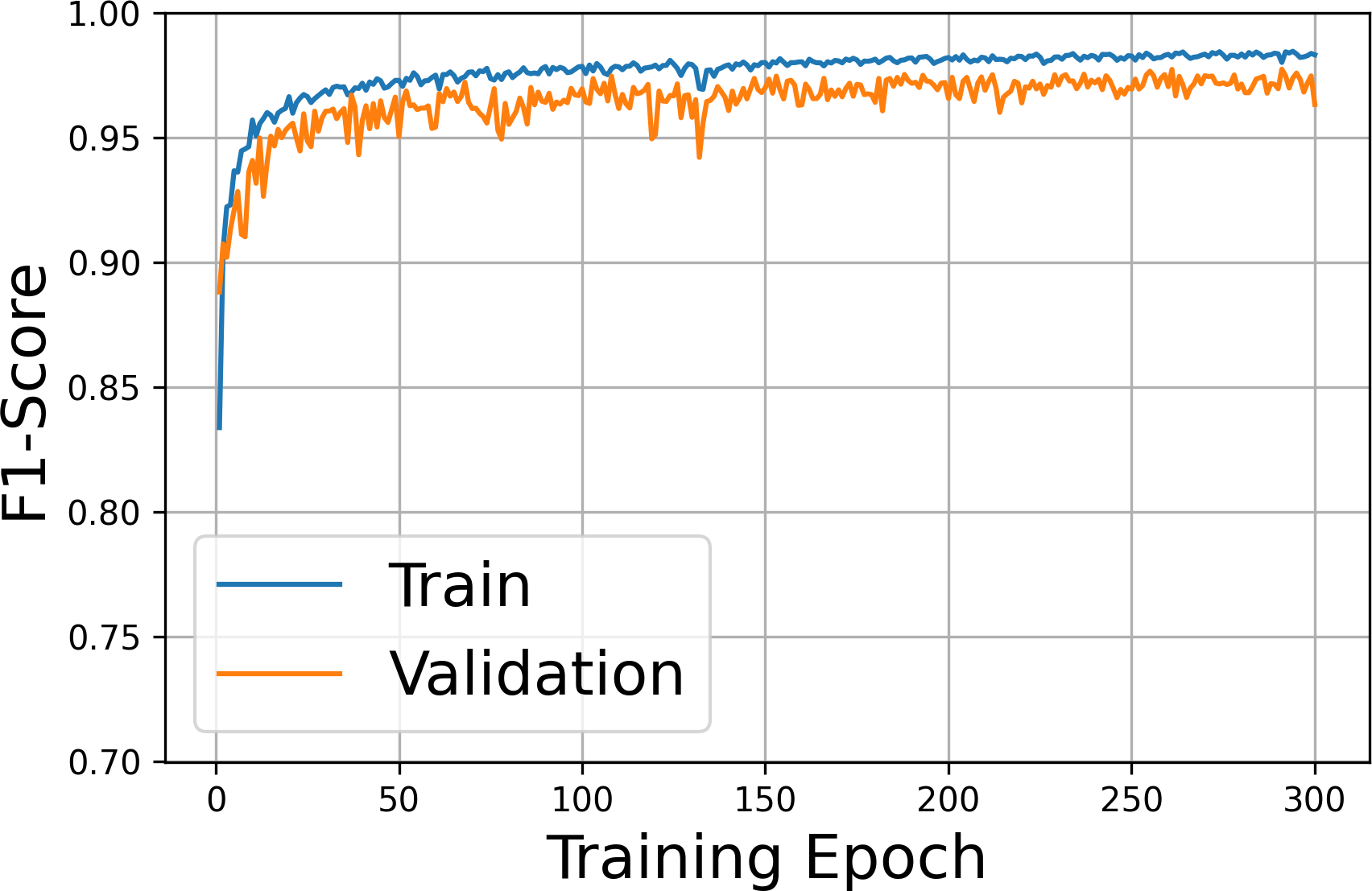}
        \caption{MSD-net MTF 1}
        \label{}
    \end{subfigure}
    %%%%%%%%%%%%%%
    \begin{subfigure}{.49\linewidth}
        \includegraphics[width=\linewidth]{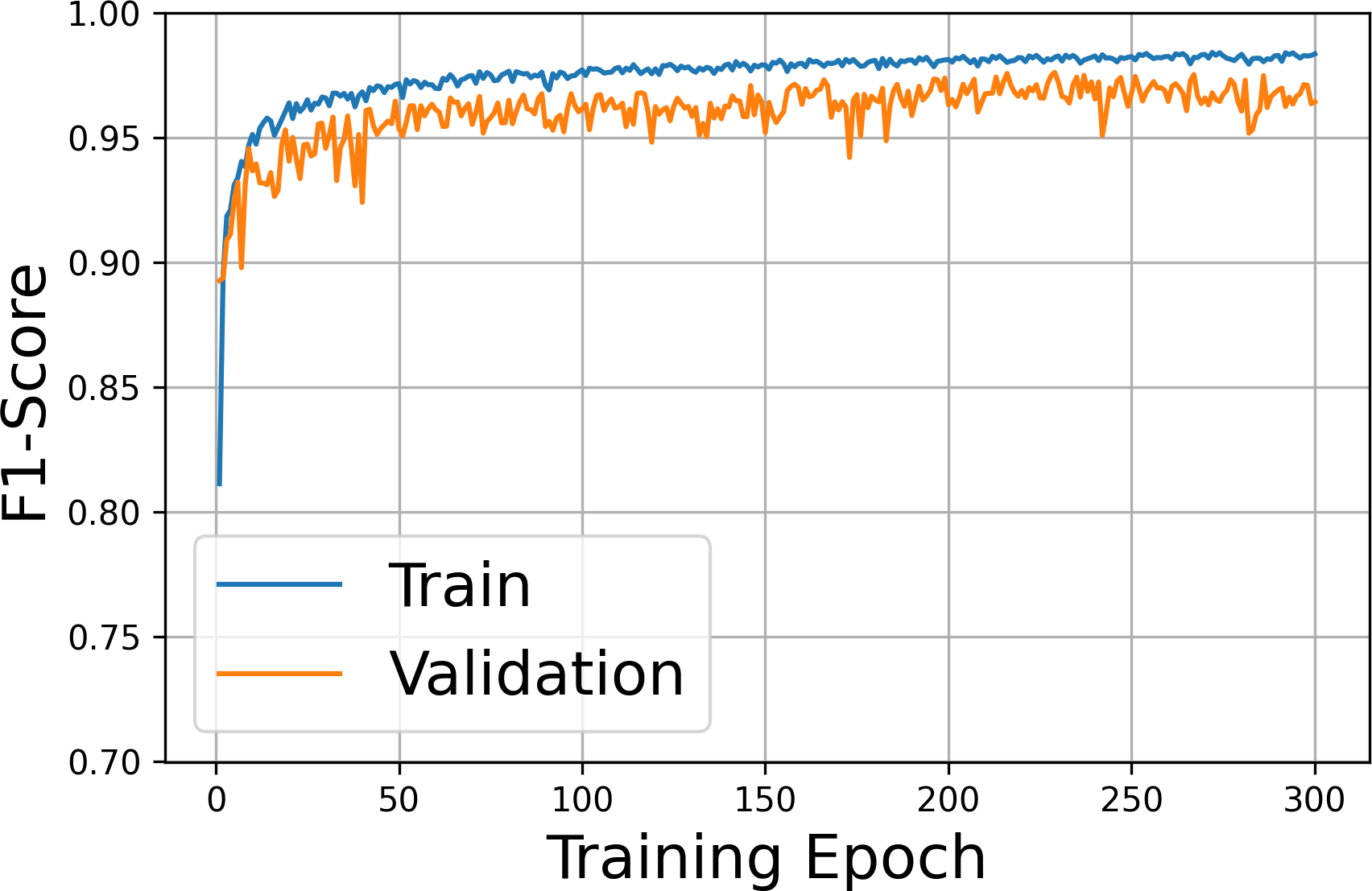}
        \caption{MSD-net MTF 2}
        \label{}
    \end{subfigure}
    %%%%%%%%%%%%%%
    \begin{subfigure}{.49\linewidth}
        \includegraphics[width=\linewidth]{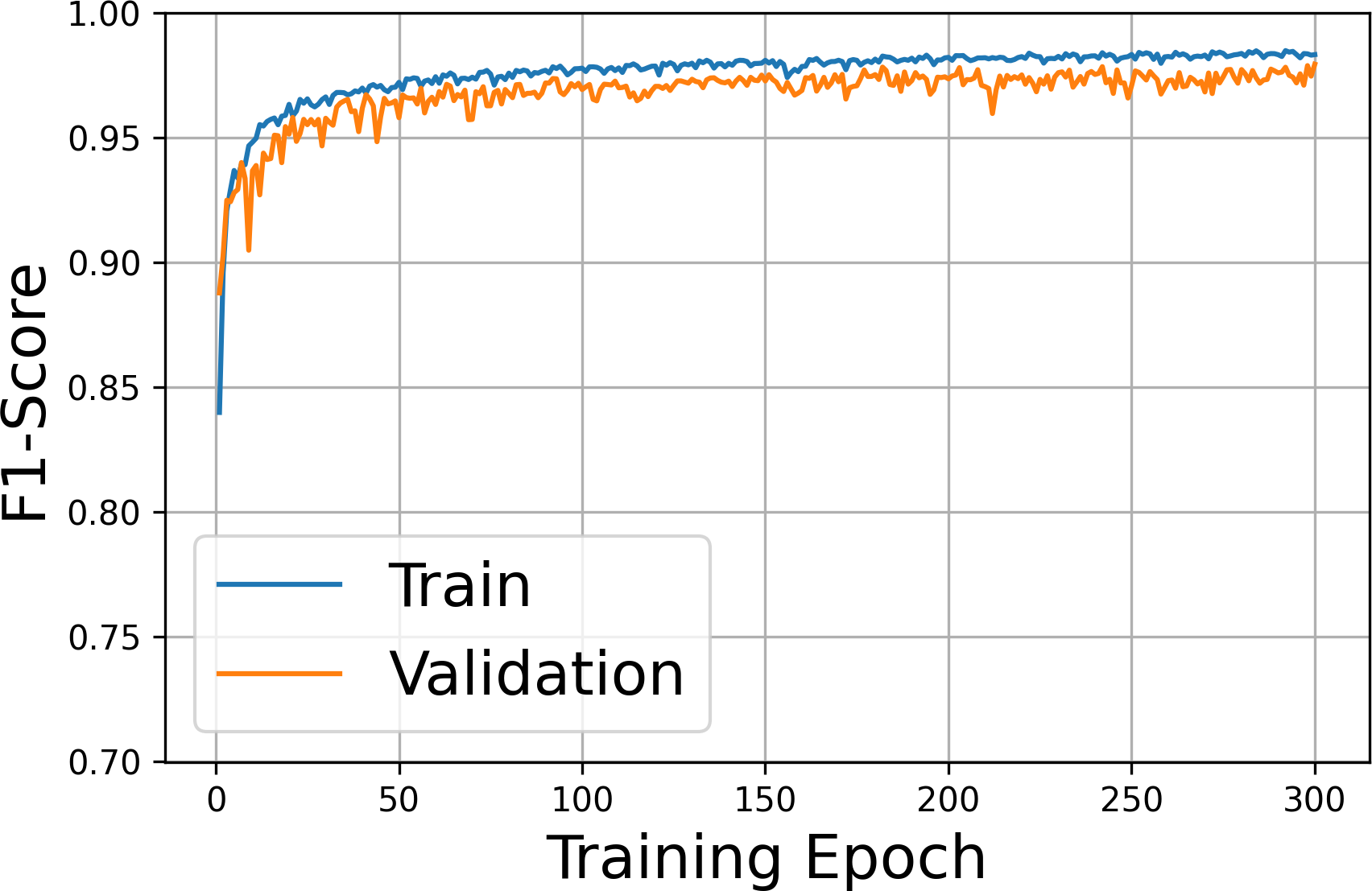}
        \caption{MSD-net MTF 3}
        \label{}
    \end{subfigure}
    %%%%%%%%%%%%%%
    \begin{subfigure}{.49\linewidth}
        \includegraphics[width=\linewidth]{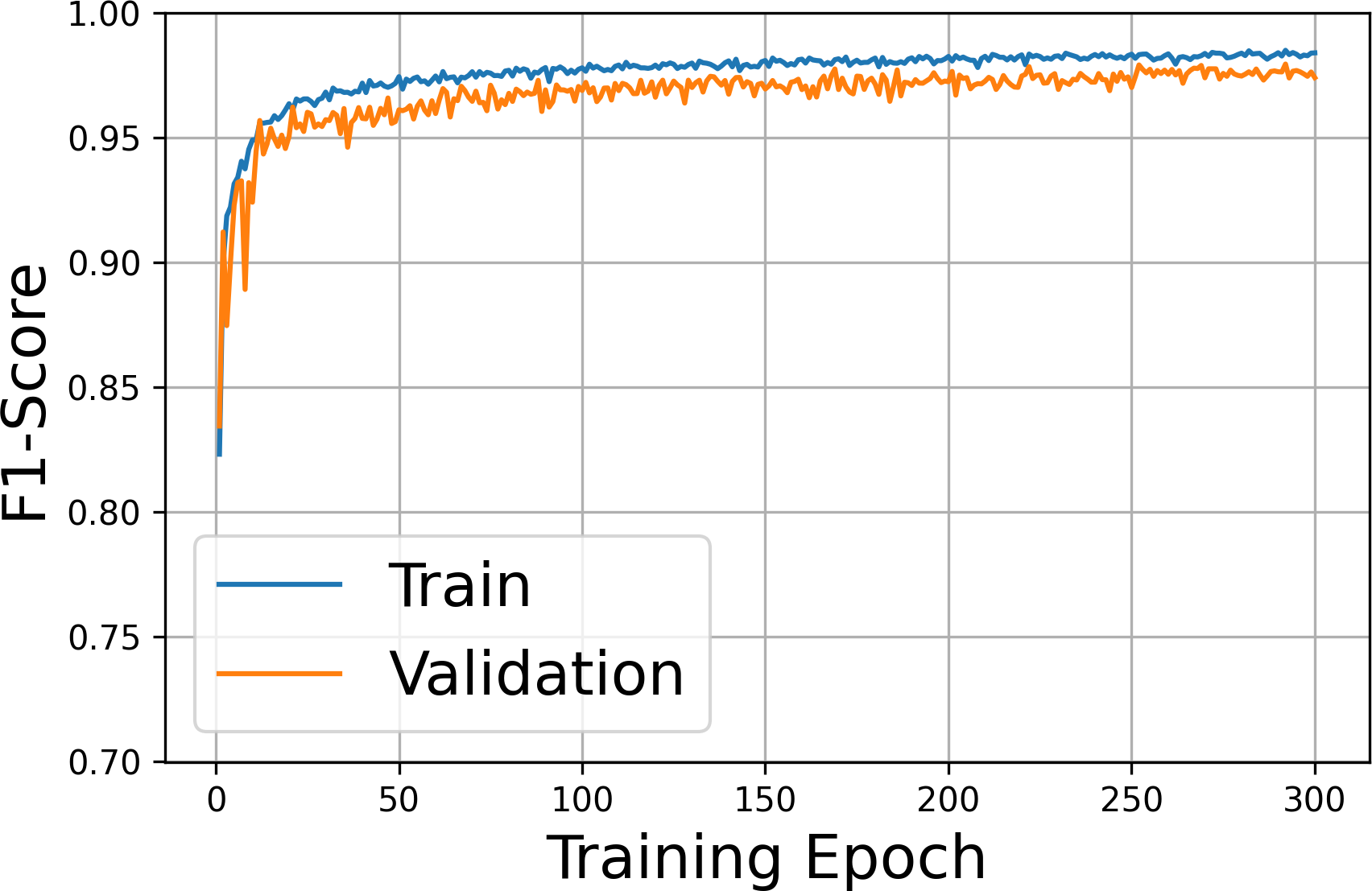}
        \caption{MSD-net MTF 4}
        \label{}
    \end{subfigure}
    %%%%%%%%%%%%%%
    \begin{subfigure}{.49\linewidth}
        \includegraphics[width=\linewidth]{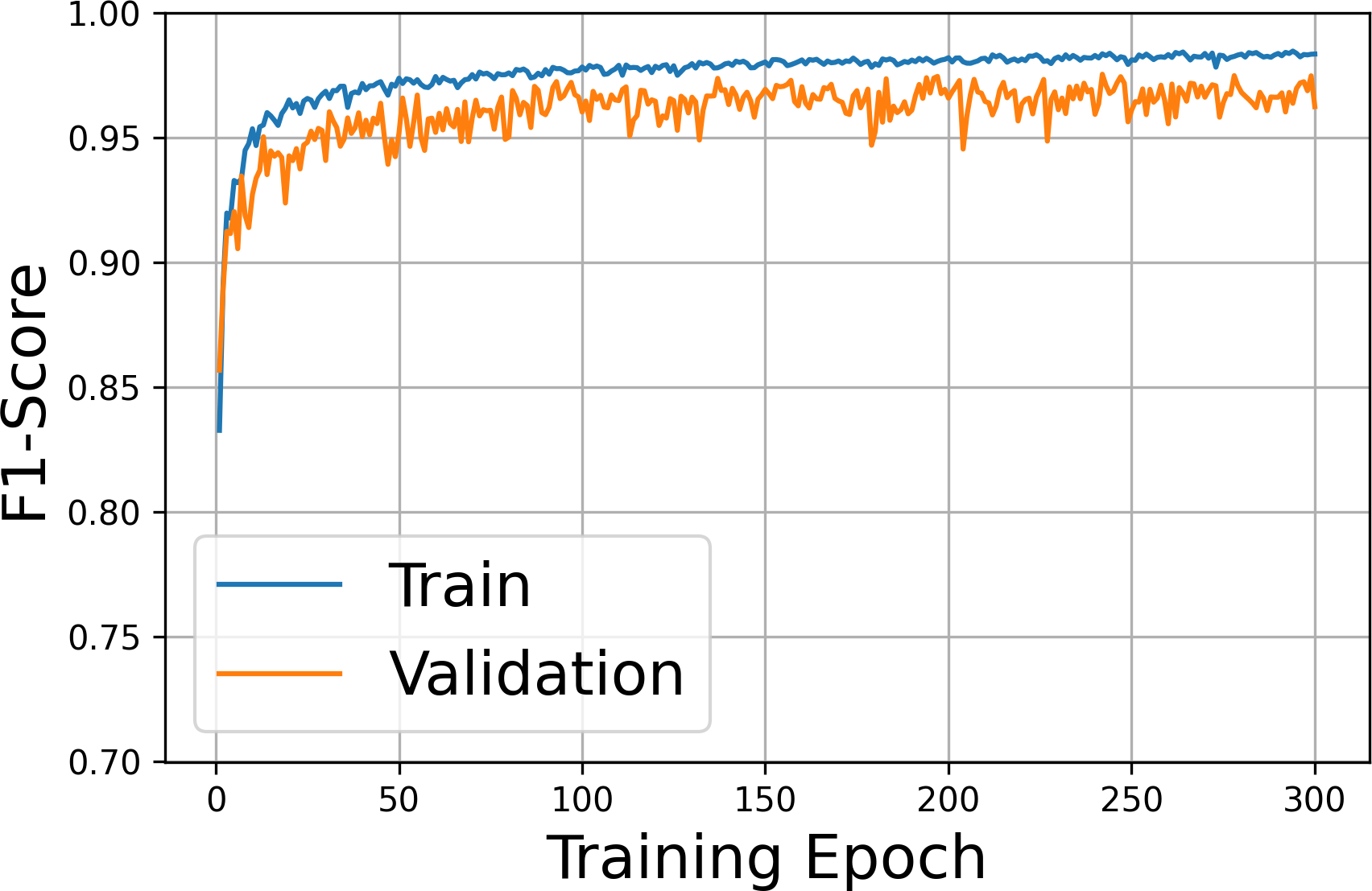}
        \caption{MSD-net MTF 5}
        \label{}
    \end{subfigure}
    %%%%%%%%%%%%%%
    \begin{subfigure}{.49\linewidth}
        \includegraphics[width=\linewidth]{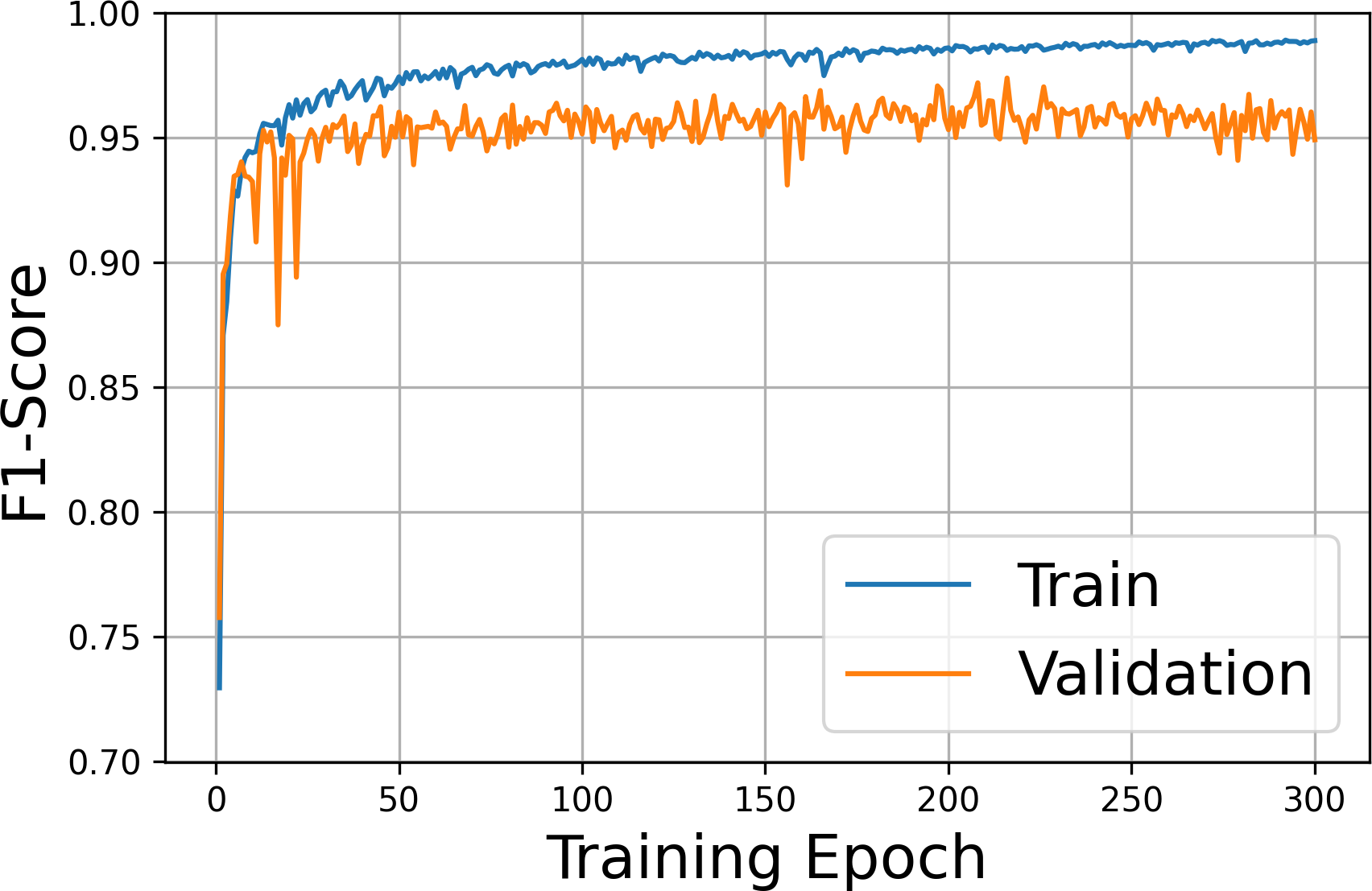}
        \caption{U-net MTF 4}
        \label{}
    \end{subfigure}
    %%%%%%%%%%%%%%
    \begin{subfigure}{.49\linewidth}
        \includegraphics[width=\linewidth]{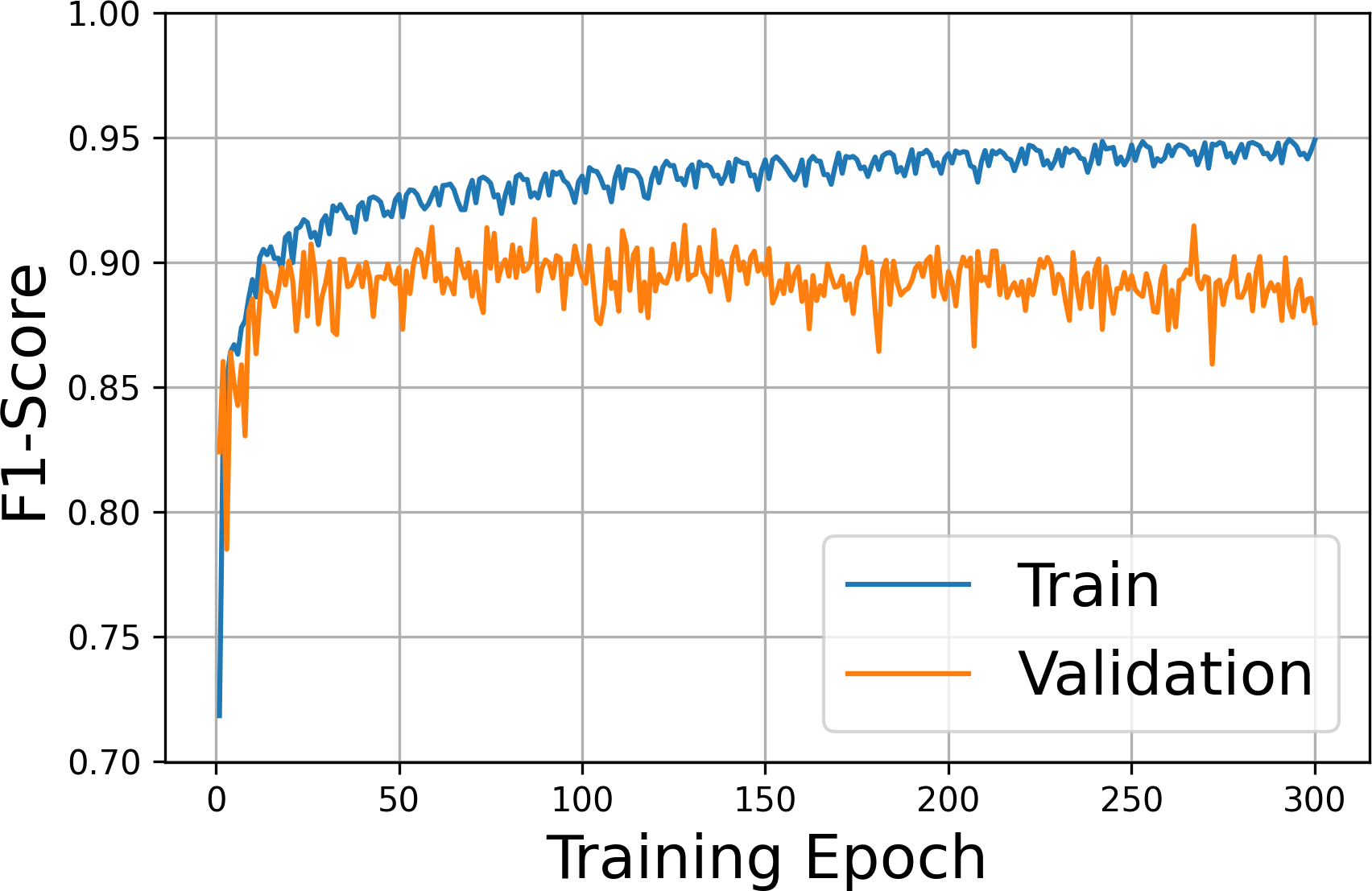}
        \caption{MSD-net MTF 4 low dose range}
        \label{}
    \end{subfigure}
    %%%%%%%%%%%%%%
    \begin{subfigure}{.49\linewidth}
        \includegraphics[width=\linewidth]{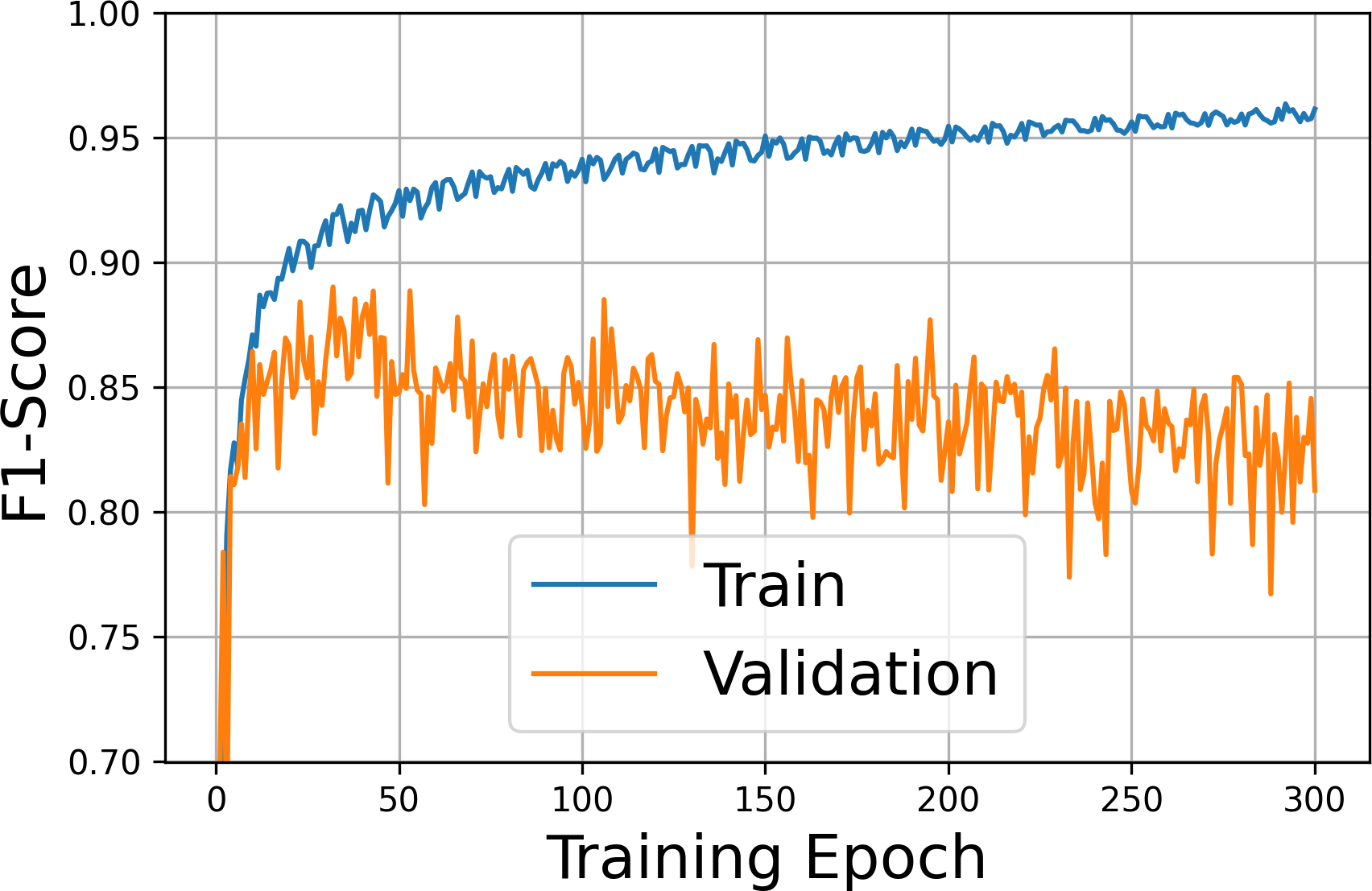}
        \caption{U-net MTF 4 low dose range}
        \label{}
    \end{subfigure}
    \caption{F1-Score learning curves for all networks applied in this work.}
    \label{fig:trainingcurves}
\end{figure}

Fig. \ref{fig:trainingcurves} shows the learning curves for all networks applied in this work. The F1-Score is shown instead of the binary cross entropy loss, as it has a more intuitive range of values, 1 being a pixel wise perfect segmentation and 0 the worst result possible. The F1-Score is a harmonic mean between the precision and recall, and properly gauges the ability to accurately classify both positives and negatives, \emph{i.e.} identify both nanoparticle and background.

\begin{figure}[H]
  \centering
  \includegraphics[width=0.98\linewidth]{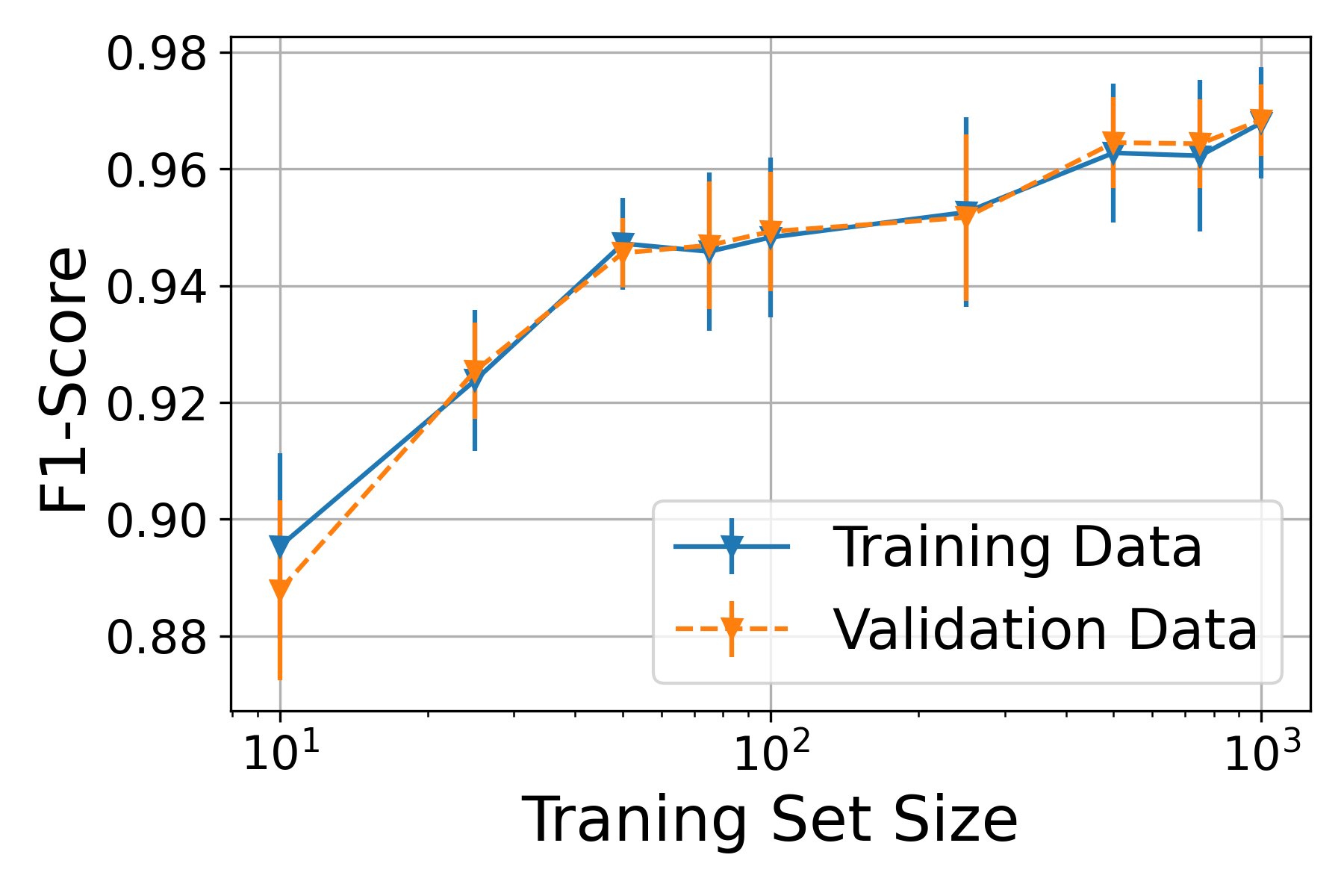}
  \caption{Average F1-Score over the MTF 4 model training dataset of 10,000 images and validation dataset of 500 images for varying sizes of training data involved in the training process. These scores are produced by an MSD-net trained on the MTF 4 model training dataset where the size of the dataset used for training is varied. The variance is used as error bars.}
  \label{fig:trainingsetsize}
\end{figure}

Fig. \ref{fig:trainingsetsize} shows the effect of varying amounts of data used for training. There is a clear improvement when going from 10 atomic structures to 100, and again from 100 to 1000, however little is gained going beyond 500 structures. This number is the number of atomic structures and number of images per training epoch. Each training epoch iterates over the 10 image epochs, which varies the microscope parameters in the image for the same set of atomic structures. Increasing the dataset size beyond 1000 structures would significantly increase training times.

Training is executed using the ADAM gradient optimiser with the AMSGrad variant activated, and a learning rate of 0.001. Many training epochs in an attempt to obtain a well converged network, which is typically the most robust when generalising to experimental data.

\section{Noise Fitting}
\label{sec:noise-fitting}

\begin{figure}[H]
  \centering
  \includegraphics[width=0.75\linewidth]{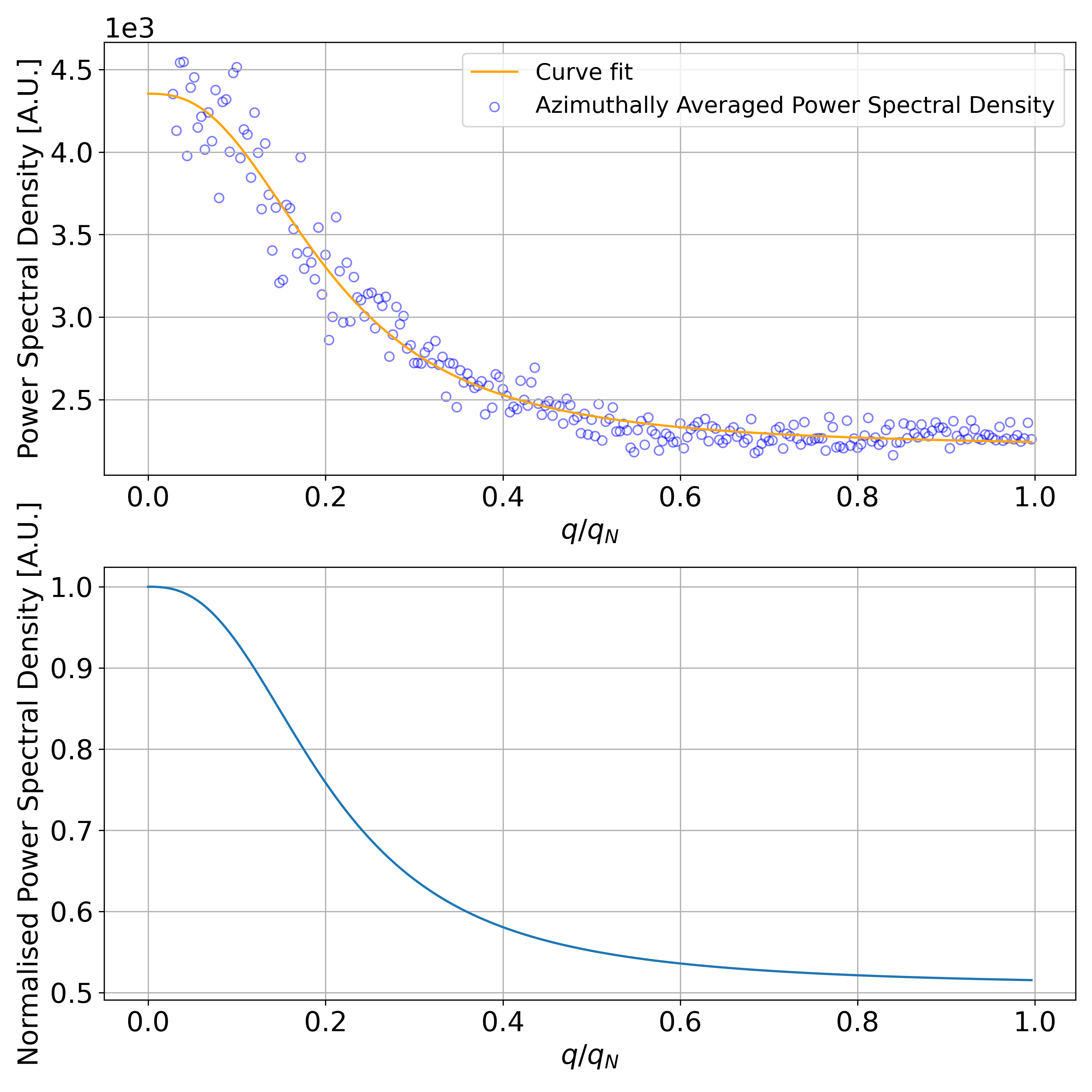}
  \caption{MTF fitted from the power spectral density of the first frame at approximately 10 $e^-/\text{\AA}^{2}$. Top: Raw fit (Eq. \ref{eq:mtf_raw}) with $R^2=0.99$, $c_0=0.20$, $c_1=4353.29$, $c_2=2207.83$, $c_3=2.57$. Bottom: Normalised MTF (Eq. 1) with $c_0=0.20$, $c_3=2.57$, $C=0.51$.}
  \label{fig:Noisefit_frame0}
\end{figure}

\begin{figure}[H]
  \centering
  \includegraphics[width=0.75\linewidth]{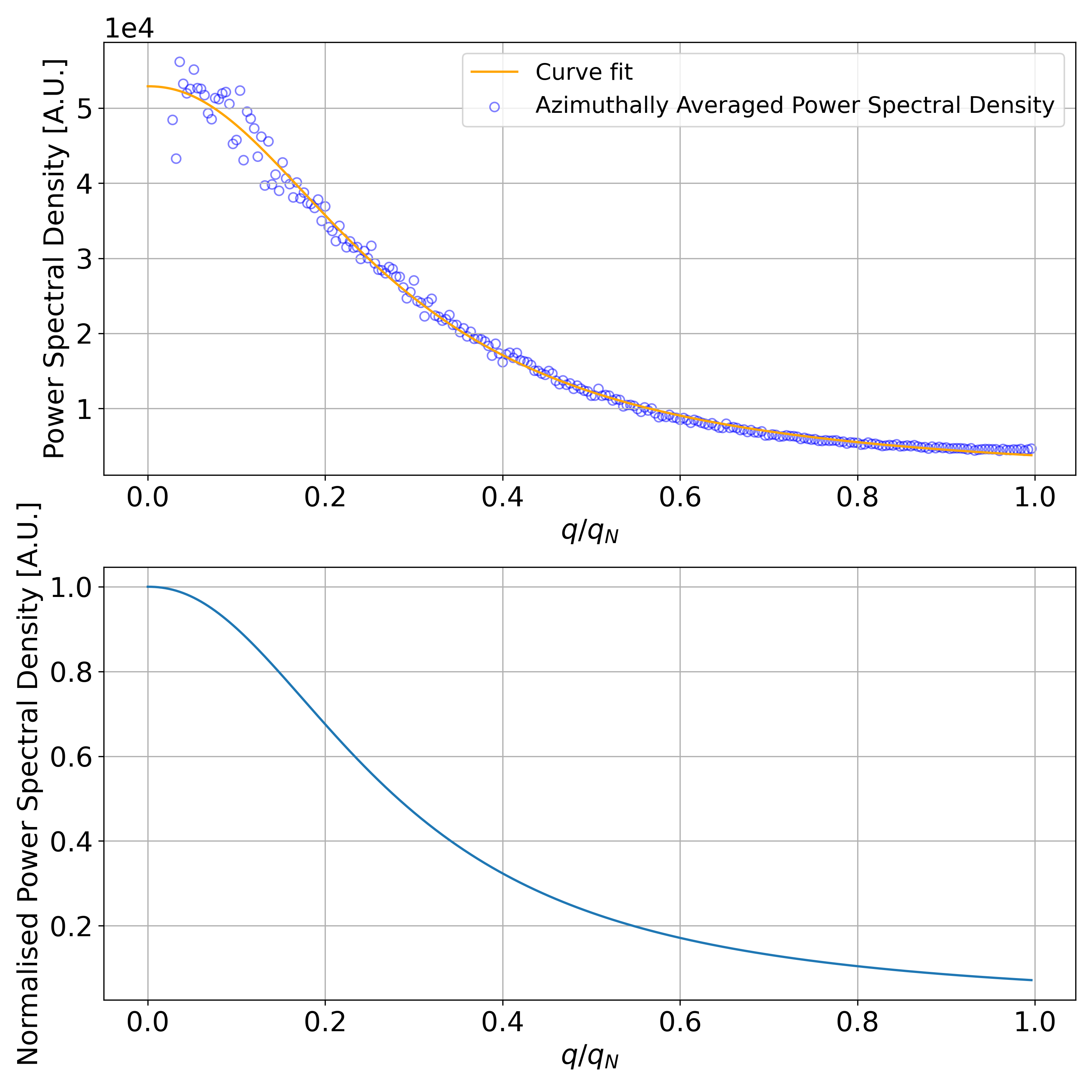}
  \caption{MTF fitted from the power spectral density of the final frame at approximately 1300 $e^-/\text{\AA}^{2}$. Top: Raw fit (Eq. \ref{eq:mtf_raw}) with $R^2=0.99$, $c_0=0.28$, $c_1=52887.32$, $c_2=602.13$, $c_3=2.25$. Bottom: Normalised MTF (Eq. \ref{eq:mtf}) with $c_0=0.28$, $c_3=2.15$, $C=0.01$.}
  \label{fig:Noisefit_framelast}
\end{figure}

The power spectral density is defined as the magnitude squared of the Fourier transform. This is taken from a vacuum region of the HR-TEM image for every frame. This 2D-spectrum is averaged azimuthally to obtain a 1D-spectrum, as done in Ref. \cite{Vulovic2010AMicroscopy}. Frequencies are normalised by the Nyquist frequency, \emph{i.e.} $\tilde{q} = q/q_N$. As shown in Figures \ref{fig:Noisefit_frame0} and \ref{fig:Noisefit_framelast}, the spatial frequency filtering affect of the MTF is clear. The raw spectrum 
\begin{equation}
    MTF(\tilde{q}) = (c_1-c_2) \cdot \frac{1}{1+(\frac{\tilde{q}}{c_0})^{c_3}} + c_2
    \label{eq:mtf_raw}
\end{equation}
is fitted in the top of the two figures. The bottom of the two figures presents the fitted MTF normalised by the $q\rightarrow0$ limit, $c_1$, \emph{i.e.} Eq. 1.

In Fig. 6, the readout noise contribution, $N_0$, is fitted from its fractional contribution the the total noise floor as a function of electron dose per pixel. A simulation of vacuum was made, where $c_0$ and $c_3$ were varied within a range of $[0.2, 0.3]$ and $[2, 3]$ respectively, and an addition Poisson noise is added with $N_0 \pm 50\%$. Fitting MTFs for the simulated image series presents the same behaviour of the $C$ parameter as in Fig. 5, proving a sensible modelling of the readout and shot noise contributions. 

\begin{figure}[H]
  \centering
  \includegraphics[width=0.75\linewidth]{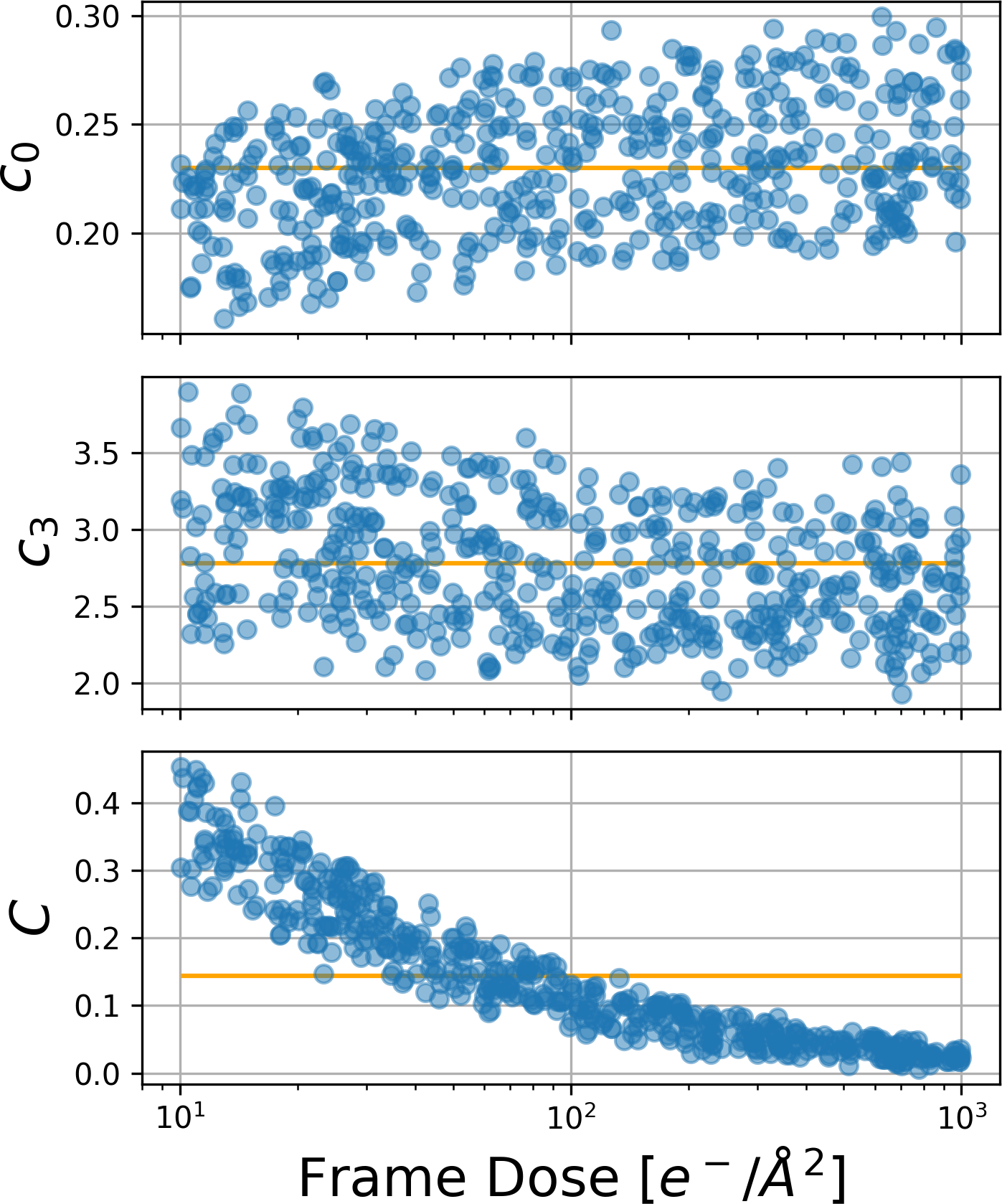}
  \caption{Distribution of parameters for Eq. 1, fitted to a simulated vacuum image series of continuously increasing dose. This replicates the behaviour in Fig. 5.}
  \label{fig:mtfvsdose_simulated}
\end{figure}

\section{Simulated MTF Comparison}
\label{sec:simulated_mtf_comparison}
\begin{figure}[H]
	\centering
	\begin{subfigure}{.49\textwidth}
		\includegraphics[width=\linewidth]{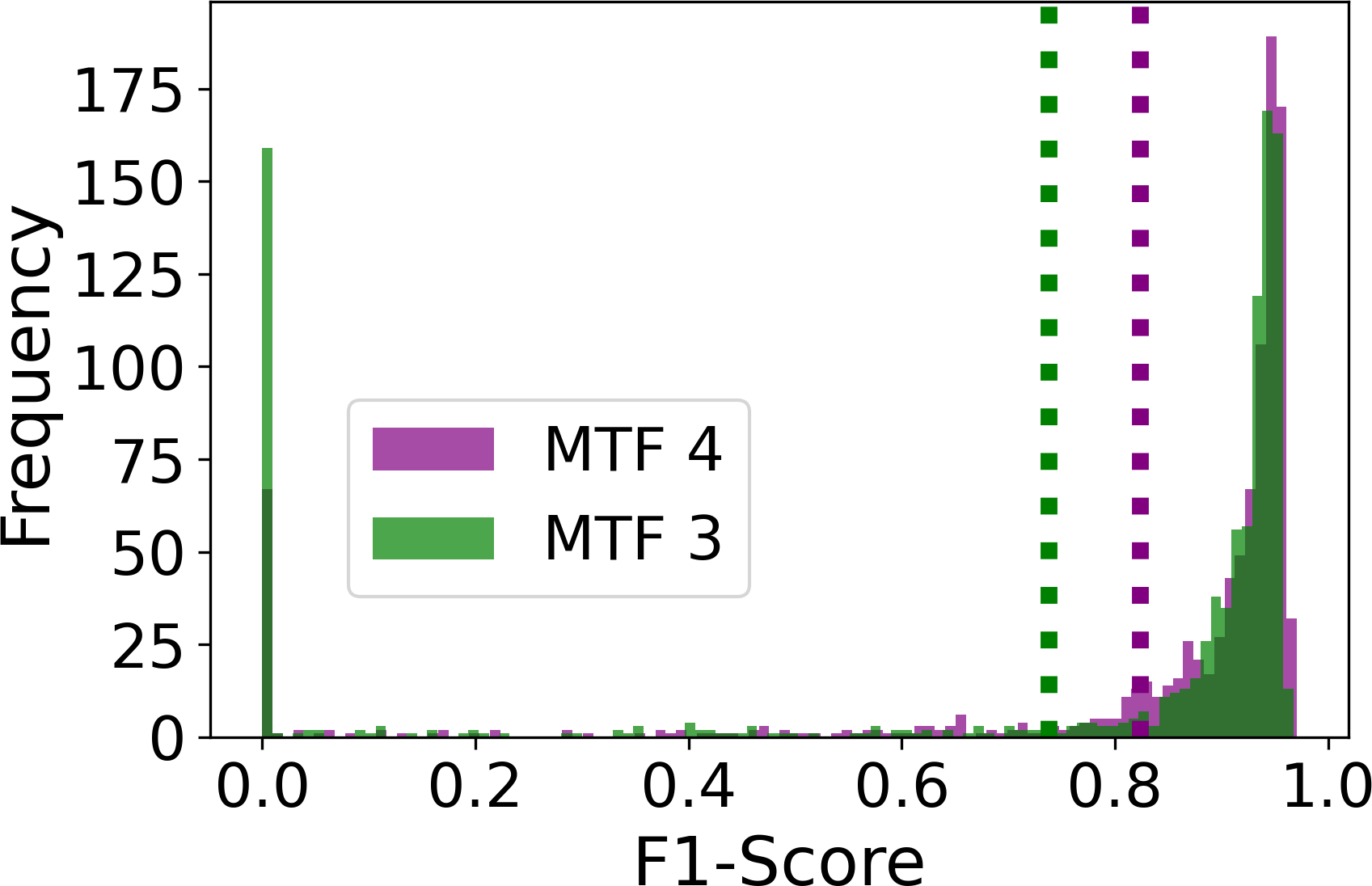}
        \caption{}
		\label{fig:simulateddataanalysis_hist}
	\end{subfigure}
%%%%%%%%%%%%%%
	\begin{subfigure}{.47\textwidth}
		\includegraphics[width=\linewidth]{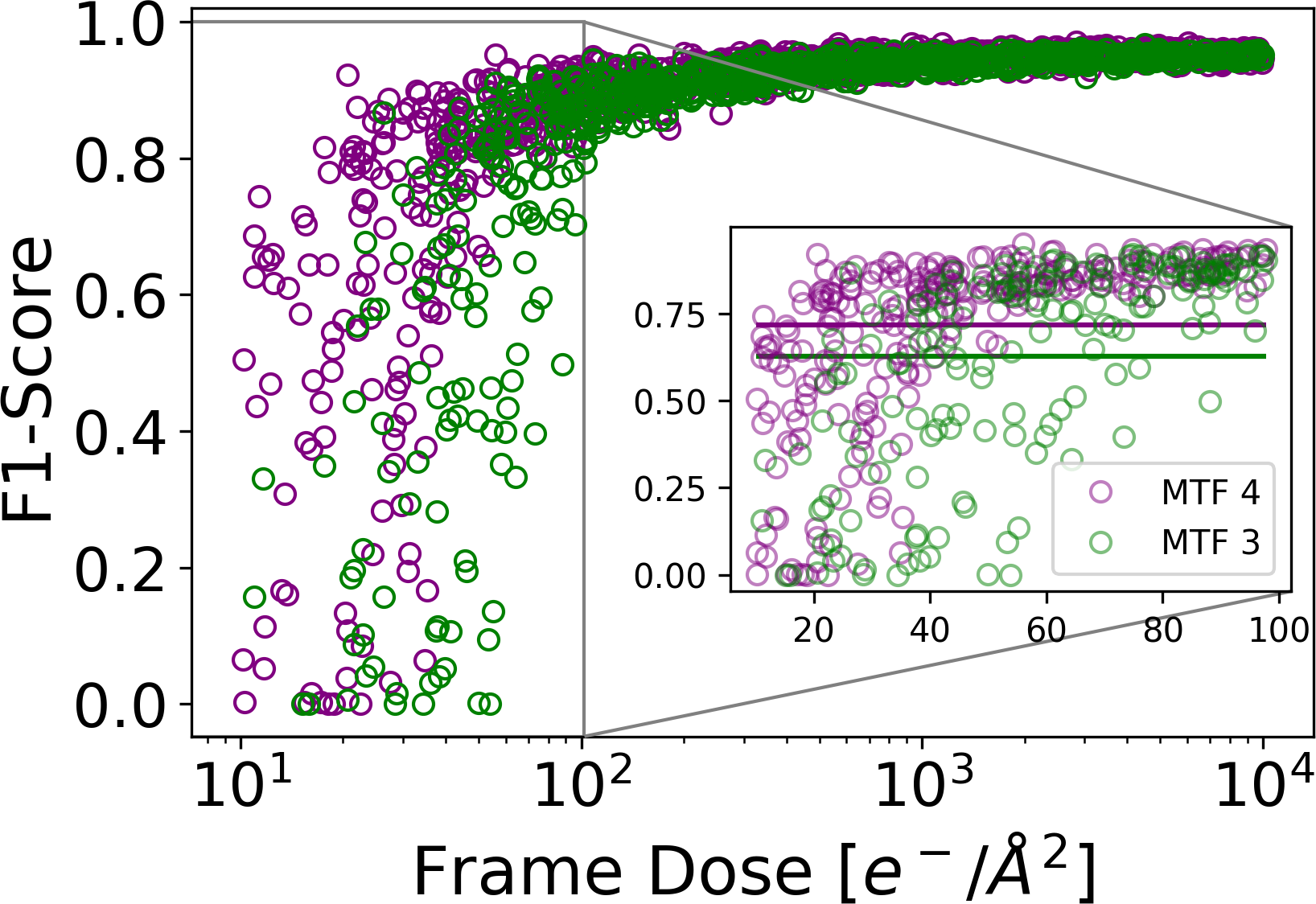}
        \caption{}
        \label{fig:simulateddataanalysis_dose}
	\end{subfigure}
	\caption{Comparing the networks trained separately on simulated data with the MTF 4 model and the MTF 3 model. Both networks are applied on a simulated continuously increasing dose image series of a CeO$_2$ supported Au nanoparticle with the MTF 3 model applied. The F1-Score ranges between 0 and 1, where 1 is a perfect segmentation. The MTF 4 model learns to distinguish noise and signal better than the MTF 3 model, seen by the higher mean F1-Score. Top: Histogram of F1-Score for each image. Dashed lines represent the colour coded mean. Bottom: F1-Score as a function of frame dose. Solid lines represent the mean within the inset of the very low dose regime.}
	\label{fig:simulateddataanalysis}
\end{figure}

\end{document}